\documentclass[preprints,article,accept,pdftex,moreauthors]{Definitions/mdpi} 

\firstpage{1} 
\pubvolume{1}
\issuenum{1}
\articlenumber{0}
\pubyear{2024}
\copyrightyear{2024}
\datereceived{ } 
\daterevised{ } 
\dateaccepted{ } 
\datepublished{ } 
\hreflink{https://doi.org/} 

\usepackage{physics}

\Title{Efficient implementation of discrete-time quantum walks on quantum computers}

\TitleCitation{Efficient implementation of discrete-time quantum walks on quantum computers}

\Author{Luca Razzoli $^{1,2}$*\orcidA{}, Gabriele Cenedese $^{1,2}$\orcidB{}, Maria Bondani$^{3}$\orcidC{}, and Giuliano Benenti $^{1,2,4}$\orcidD{}}

\AuthorNames{Luca Razzoli, Gabriele Cenedese, Maria Bondani, and Giuliano Benenti}

\AuthorCitation{Razzoli, L.; Cenedese, G.; Bondani, M; Benenti, G.}

\address{%
$^{1}$ \quad Center for Nonlinear and Complex Systems, Dipartimento di Scienza e Alta Tecnologia, Universit\`a degli Studi dell'Insubria, via Valleggio 11, 22100 Como, Italy\\
$^{2}$ \quad Istituto Nazionale di Fisica Nucleare, Sezione di Milano, via Celoria 16, 20133 Milano, Italy\\
$^{3}$ \quad Istituto di Fotonica e Nanotecnologie, Consiglio Nazionale delle Ricerche, Via Valleggio 11, 22100 Como, Italy\\
$^{4}$ \quad NEST, Istituto Nanoscienze-CNR, P.zza San Silvestro 12, 56127 Pisa, Italy}

\corres{Correspondence: luca.razzoli@uninsubria.it}

\abstract{Quantum walks have proven to be a universal model for quantum computation and to provide speed-up in certain quantum algorithms. The discrete-time quantum walk (DTQW) model, among others, is one of the most suitable candidates for circuit implementation, due to its discrete nature. Current implementations, however, are usually characterized by quantum circuits of large size and depth, which leads to a higher computational cost and severely limits the number of time steps that can be reliably implemented on current quantum computers. In this work, we propose an efficient and scalable quantum circuit implementing the DTQW on the $2^n$-cycle based on the diagonalization of the conditional shift operator. For $t$ time-steps of the DTQW, the proposed circuit requires only $O(n^2 + nt)$ two-qubit gates compared to the $O(n^2 t)$ of the current most efficient implementation based on quantum Fourier transforms. We test the proposed circuit on an IBM quantum device for a Hadamard DTQW on the $4$- and $8$-cycle characterized by periodic dynamics and recurrent generation of maximally entangled single-particle states. Experimental results are meaningful well beyond the regime of few time steps, paving the way for reliable implementation and use on quantum computers.}


\begin{document}

\section{Introduction}
Quantum walks were formally introduced in the 1990s as the quantum analogue of classical random walks \cite{PhysRevA.48.1687,PhysRevA.58.915}, but the seminal concept dates back to Feynman's checkerboard (see Problem 2-6 in \cite{Feynman_QWpath}), which connects the spin with the zig-zag propagation of a particle spreading at the speed of light across a (1+1)-dimensional spacetime lattice. Since their introduction, it was immediately clear the great potential of exploiting the peculiar features of quantum walks---quantum superposition of multiple paths, ballistic spread (faster than the diffusive spread of a classical random walker), and entanglement---for algorithmic purposes \cite{ambainis2003qwalg,childs2003exponential,kendon2006rw}.
Nowadays, quantum walks have proven to be a universal model for quantum computation \cite{PhysRevA.81.042330,Singh2021,Chawla2023,PhysRevLett.102.180501,Lahini2018} and examples of their use include algorithms for quantum search \cite{PhysRevA.67.052307,PhysRevA.70.022314,PhysRevLett.124.180501,PhysRevLett.129.160502}, solving hard $K$-SAT instances \cite{Campos2021}, graph isomorphism problems \cite{Douglas_2008,Tamascelli_2014,Schofield2020}, algorithms in complex networks \cite{Loke2016,Chawla2020}---such as link prediction \cite{PhysRevA.107.032605,Goldsmith_2023} and community detection \cite{PhysRevX.4.041012,PhysRevResearch.2.023378}---, and quantum simulation \cite{berry2012blackbox,PhysRevA.81.062340,PhysRevA.88.042301,Arrighi2016,Molfetta_2016}.
Furthermore, quantum communication protocols based on quantum walks have been put forward \cite{PhysRevA.90.012331,Yalçınkaya_2015,Wang2017,Shang_2018,Srikara2020,PhysRevA.107.022611,bottarelli2023qrouting,vlachou2018,AbdEl-Latif2020}.
For comprehensive reviews on quantum walks and their applications, we refer the reader to \cite{Venegas_Andraca2012,portugal2018quantum,kadian2021review} and to \cite{wang2013physical,Gräfe_2016,Neves_2018} for the their physical implementations.

There are two main models of quantum walk: discrete-time quantum walk (DTQW) \cite{PhysRevA.48.1687} and continuous-time quantum walk (CTQW) \cite{PhysRevA.58.915}. The CTQW is defined on the position Hilbert space of the quantum walker and the evolution is driven by the Hamiltonian $H$ of the system, $U(t) \equiv \exp(-i t H /\hbar)$. The DTQW is defined on a Hilbert space comprising an additional coin space and the evolution is driven by a position shift operator, $S$, controlled by a quantum coin operator, $C$, acting at discrete time steps. The single time-step operator is defined as $U=S(C\otimes I_p)$, with $I_p$ the identity in position space, from which $U(t) \equiv U^t$ with $t \in \mathbb{N}$. 
As pointed out in  \cite{Loke_2017}, implementing the evolution operator of a DTQW is simplified by the fact that (i) the time is discrete, (ii) the evolution is repetitive, $U(t)=U^t$, and (iii) $U$ acts locally on the coin-vertex states encoding the graph (applying $U$ to the coin-vertex states associated to a given vertex will propagate the corresponding amplitudes only to adjacent vertices).

DTQWs already have efficient physical implementations in platforms that natively support the conditional walk operations, e.g., photonic systems \cite{Gräfe_2016,Neves_2018}. However, devising efficient implementations on digitized quantum computers is desirable and necessary in order to make DTQWs available to develop quantum algorithms for general purpose quantum computers and, in general, quantum protocols to be implemented in circuit models. A first circuit implementation of a DTQW on the cycle was realized on a multiqubit nuclear-magnetic-resonance system  \cite{PhysRevA.72.062317}, and thereafter proposals of efficient implementation on certain graphs \cite{PhysRevA.79.052335,PhysRevA.80.062301,PhysRevA.86.042338,Wing-Bocanegra2023}, for  position-dependent coin operators \cite{Nzongani2023}, and for staggered quantum walks (a 
coinless discrete-time model of quantum walk) \cite{Acasiete2020} have been devised.

In this work we propose an efficient and scalable quantum circuit implementing the DTQW on the $2^n$-cycle, a finite discrete line with $2^n$ vertices and periodic boundary conditions. Although this is the simplest DTQW one may think of, implementing it on quantum computers already highlights the limitations of actual quantum devices \cite{slimen2021,olivieri2021,PhysRevA.103.022408,wadhia2023cycle}. In this model, the position state of the quantum walker is encoded in a $n$-qubit state. To the best of our knowledge, the most efficient state-of-the-art implementation of a DTQW \cite{Shakeel2020} overall requires $O(n^2 t)$ two-qubit gates for $t$ time steps, because it involves one quantum Fourier transform (QFT) and one inverse  QFT (IQFT) at \textit{each} time step. In quantum computers, two-qubit gates are the noisiest and take the longest time to execute, so any efficient quantum circuit should aim at significantly reducing their number. Our quantum circuit accomplishes this task through a wise use of the unitary property of the QFT: Independently of $t$, our circuit involves only one QFT (at the beginning) and one IQFT (at the end), and thus it overall requires $O(n^2 + nt)$ two-qubit gates. Accordingly, the advantage gets larger and larger for long times, passing, for $t \gg n$, from $O(n^2 t)$ in \cite{Shakeel2020} to $O(n t)$ in the present scheme. For illustrative purposes, we implemented the proposed quantum circuit on an actual quantum hardware---\texttt{ibm\_cairo}, a 27-qubit high-fidelity quantum computer---considering a Hadamard DTQW on the $4$- and $8$-cycle. Results indicate that our circuit outperforms current efficient circuits also in the regime of few time steps and provide experimental evidence of the recurrent generation of maximally entangled single-particle states in the 4-cycle \cite{PhysRevA.108.L020401}. 

The paper is organized as follows. Sec. \ref{sec:dtqw_cycle} reviews the DTQW model on the $N$-cycle. Sec.  \ref{sec:proposed_qc_dtqw} introduces the efficient quantum circuit we designed for the DTQW on the $2^n$-cycle and compares it with other existing schemes. Sec. \ref{sec:results} presents and discusses the results from testing the proposed circuit on a quantum hardware. Finally, Sec.~\ref{sec:conclusions} is devoted to conclusions and perspectives. Selected technical details are deferred to the appendices.

\section[The model: DTQW on the N-cycle]{The model: DTQW on the $N$-cycle}
\label{sec:dtqw_cycle}
\begin{figure}[t]
\includegraphics[width=0.9\textwidth]{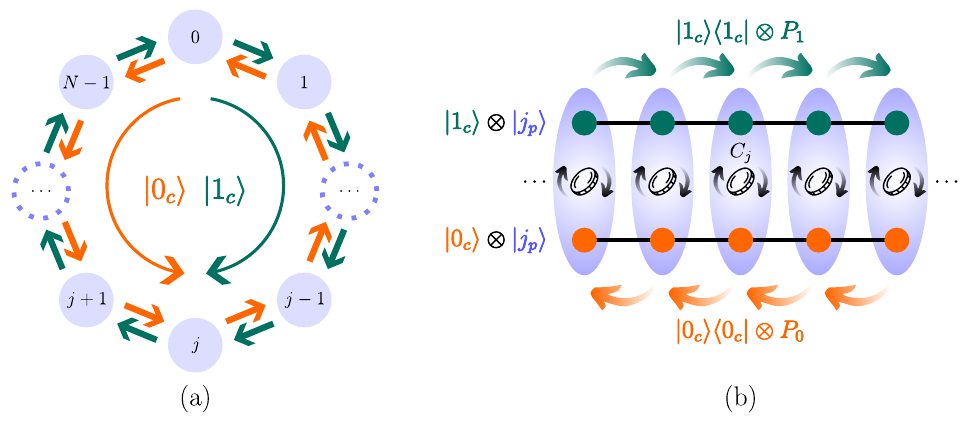}
\caption{Schematic representation of a DTQW on the $N$-cycle.
(\textbf{a}) The coin state (internal degree of freedom) is responsible for making the walker move in the cycle clockwise if $\vert 1_c \rangle$ and counterclockwise if $\vert 0_c \rangle$.   
(\textbf{b}) States and operators of a DTQW. The vertices of the cycle (light violet)---walker's position states---are labeled by $\vert j_p \rangle$ with $j=0,1,\ldots,N-1$, and each vertex comprises two sub-vertices---coin states---labeled by $\vert 0_c \rangle$ (orange) and $\vert 1_c \rangle$ (green). Each step of the walk, Eq. \eqref{eq:dtqw_step}, involves the action of a local coin operator $C_j$ responsible for mixing the coin states of each vertex (we assume $C_j = C$ $\forall j$), followed by the action of the conditional-shift operators $\vert 0_c \rangle\langle 0_c \vert \otimes P_0$ (decrement) and $\vert 1_c \rangle\langle 1_c \vert \otimes P_1$ (increment) responsible for shifting the position states, see Eq. \eqref{eq:S_comput_basis} \cite{wang2013physical}.
\label{fig:dtqw_sketch}}
\end{figure}

A $N$-cycle, or circle, is a 1D lattice having $N$ vertices and periodic boundary conditions. To each vertex, labeled by $j=0,\ldots,N-1$, we associate a quantum state, $\vert j \rangle$, which represents the walker localized at such vertex. In a DTQW, the quantum walker has an \textit{external} degree of freedom, the position, and an \textit{internal} one, the coin.
Associated to each degree of freedom is a Hilbert space: A $N$-dimensional position Hilbert space $\mathcal{H}_p^{(N)}=\operatorname{span}\left( \{\vert j_p \rangle : j=0,\ldots,N-1 \}\right)$ and a two-dimensional coin Hilbert space $\mathcal{H}_c^{(2)}=\operatorname{span}\left( \{\vert s_c \rangle : s=0,1 \}\right)$.  We use the label ``p'' to refer to walker's position degree of freedom and ``c'' to coin degree of freedom. Depending on the coin state, the walker can move counterclockwise ($s=0$) or clockwise ($s=1$) on the cycle (Fig. \ref{fig:dtqw_sketch}).
The full Hilbert space $\mathcal{H} \equiv \mathcal{H}_c^{(2)} \otimes \mathcal{H}_p^{(N)}$ is
\begin{equation}
\mathcal{H} = \operatorname{span} \left( \{\vert s_c \rangle \vert j_p \rangle : s=0,1 ; j=0,\ldots,N-1 \} \right). 
\label{eq:extended_position_basis}
\end{equation}

\noindent This is the natural basis for a DTQW and in the following we will refer to it as the computational basis. The coin basis state are $\vert 0_c \rangle = (1,0)^\intercal$ and $\vert 1_c \rangle= (0,1)^\intercal$, with ${}^\intercal$ denoting the transpose without complex conjugation; the position basis states are $\vert j_p \rangle = (0, \ldots, 0,1,0,\ldots,0)^\intercal$, with the only nonzero element in position $j$. Accordingly, a generic coin-position basis state $\vert s_c \rangle \vert j_p \rangle=\vert s_c \rangle \otimes \vert j_p \rangle$ is represented by the column vector of length $2N$, whose first $N$ entries are related to $s=0$ and the last $N$ to $s=1$. The only nonzero entry is the $(Ns+j)$-th one.

\begin{figure}[t]
\includegraphics[width=\textwidth]{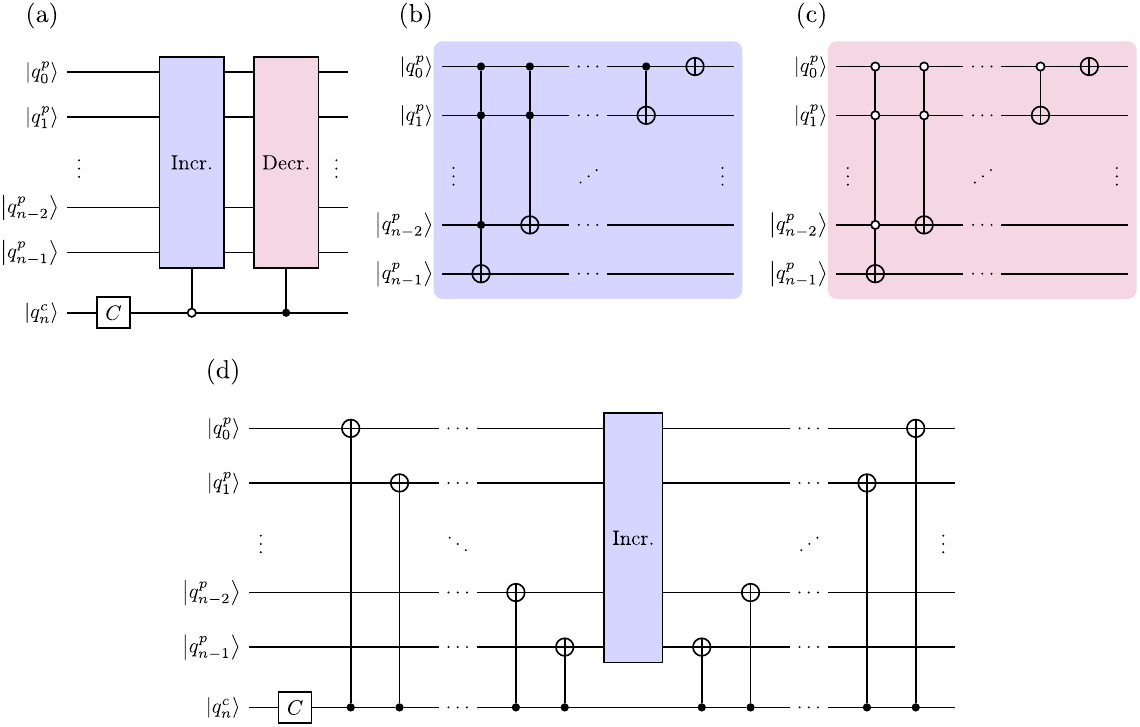}
\caption{(\textbf{a}) Quantum circuit implementing one time-step of the DTQW on the $2^n$-cycle based on controlled-\textit{increment}  (I) and controlled-\textit{decrement} (D) gates \cite{PhysRevA.79.052335}. (\textbf{b}) Increment and (\textbf{c}) decrement gates consist of generalized CNOT gates, with controls being $\vert 1 \rangle$ (solid circle) and $\vert 0 \rangle$ (empty circle), respectively. These gates act on the walker's position quantum register, conditional on the coin's qubit state [see panel (a)]. (\textbf{d}) The ID-quantum circuit shown in panel (a) can be conveniently re-designed in terms of one increment gate (not controlled by the coin qubit) and CNOT gates only, being ${\rm Decr.} = \left( \bigotimes_{k=1}^n X_k \right) {\rm Incr.} \left( \bigotimes_{k=1}^n X_k \right)$  \cite{Shakeel2020}. Quantum circuits in panels (a,d) implement the conditional shift operator $S = \sum_{j=0}^{N-1} (\vert 0_c \rangle \langle 0_c \vert \otimes \vert ( j+1 \bmod N )_p \rangle\langle j_p \vert + \vert 1_c \rangle \langle 1_c \vert \otimes \vert ( j-1 \bmod N )_p \rangle\langle j_p \vert)$ having the opposite convention to \eqref{eq:S_comput_basis} used in the present work.
\label{fig:ID_qc}}
\end{figure}

The evolution is ruled by the unitary single time-step operator
\begin{equation}
U = S (C \otimes I_p),
\label{eq:dtqw_step}
\end{equation}

\noindent with $I_p$ the identity in position space and $C$ the coin operator acting on the coin state. Coin and conditional shift operators must be unitary for $U$ to be unitary. The conditional shift operator $S$ acts on the full Hilbert space and makes the walker move according to the coin state: $S \vert s_c \rangle \vert j_p \rangle = \vert s_c \rangle \vert [(j+2s-1) \bmod N]_p \rangle$, where operations in position space are performed modulo $N$. Such operator can be written as
\begin{equation}
S = \sum_{s=0}^1 \sum_{j=0}^{N-1} \vert s_c \rangle \langle s_c \vert \otimes \vert [(j+2s-1) \bmod N]_p \rangle\langle j_p \vert \equiv \sum_{s=0}^1 \vert s_c \rangle \langle s_c \vert \otimes P_{s},
\label{eq:S_comput_basis}
\end{equation}

\noindent where $P_s \vert j_p \rangle = \vert [(j +2s-1) \bmod N]_p \rangle$. The operators $P_0$ (decrement) and $P_1$ (increment) are responsible for making the walker move one step counterclockwise and clockwise, respectively [see Fig. \ref{fig:dtqw_sketch} and Fig. \ref{fig:ID_qc}(b)-(c)]. In the computational basis the conditional shift operator \eqref{eq:S_comput_basis}  has matrix representation
\begin{equation}
S =
\begin{pmatrix}
P_0 & \mathbf{0}\\
\mathbf{0} & P_1
\end{pmatrix},
\label{eq:S_matrix}
\end{equation}

\noindent where $\mathbf{0}$ is the $N \times N$ null matrix and
\begin{equation}
P_0 =
\begin{pmatrix}
0 		& 1 & 0 & \ldots & 0\\
0 		& \ddots & \ddots & \ddots & \vdots\\
0  		& \ddots &  \ddots & 1 & 0\\
\vdots  & \ddots &  0 & 0 & 1\\
1 		& \cdots & 0 & 0 & 0
\end{pmatrix}, 
\qquad
P_1 =
\begin{pmatrix}
0 		& 0 & 0 & \ldots & 1\\
1 		& \ddots & \ddots & \ddots & \vdots\\
0  		& \ddots &  \ddots & 0 & 0\\
\vdots  & \ddots &  1 & 0 & 0\\
0 		& \cdots & 0 & 1 & 0
\end{pmatrix} = P_0^\intercal .
\label{eq:P0_P1_matrix}
\end{equation}

The quantum walker is usually assumed to be initially localized at the vertex $\vert 0 \rangle$,
while the coin is in a generic superposition state,
\begin{equation}
\vert \psi_0 \rangle = \left[ \cos \left(\frac{\theta}{2} \right) \vert 0_c \rangle + e^{i \phi}\sin \left(\frac{\theta}{2} \right) \vert 1_c \rangle \right] \vert 0_p \rangle,
\label{eq:psi0}
\end{equation}

\noindent with $\theta \in [0,\pi]$ and $\phi \in [0,2\pi[$. After $t\in \mathbb{N}$ time steps the quantum walker will be in the state
\begin{equation}
\vert \psi_t \rangle  = U^t \vert \psi_0 \rangle = \sum_{j=0}^{N-1} \left[ \psi_{0,j}(t) \vert 0_c \rangle \vert j_p \rangle + \psi_{1,j}(t) \vert 1_c \rangle \vert j_p \rangle \right],
\label{eq:psit}
\end{equation}

\noindent where the amplitudes $\psi_{s,j}(t)\in \mathbb{C}$---with $s=0,1$---associated to the states $\vert s_c \rangle \vert j_p \rangle$ satisfy the normalization condition at any $t$, $\sum_{j=0}^{N-1}\sum_{s=0}^{1} \vert \psi_{s,j}(t) \vert^2 =1 $. The probability to find the walker at position $k$ at time $t$, irrespective of the coin state, is $p_k(t)= \sum_{s=0}^{1} \vert \psi_{s,k}(t) \vert^2$.

\section[Quantum circuit implementing the DTQW on the 2\^n-cycle]{Quantum circuit implementing the DTQW on the $2^n$-cycle}
\label{sec:proposed_qc_dtqw}
A quantum circuit implementing the DTQW on the $N$-cycle with $N=2^n$ requires $n+1$ qubits: $n$ to encode the walker's position state and an additional $1$ to encode the coin state. Both states are encoded in base 2. Denoting by $j_2$ the binary representation of the integer $j$ with $n$ digits, in the \textit{little-endian} ordering convention (the most significant bit is placed on the left) we write
\begin{equation}
\vert j \rangle \equiv \vert j_2\rangle = \vert q_{n-1}\ldots q_0\rangle,
\end{equation}

\noindent where $q_k = 0,1$ with $k = 0,\ldots,n-1$, such that $j = \sum_{k=0}^{n-1} q_k \times 2^k$. Accordingly, we write the quantum state of the quantum walker as the $(n+1)$-qubit state 
\begin{equation}
\vert s_c \rangle \vert j_p \rangle \equiv \vert q_n^{c} q_{n-1}^{p} \ldots q_{0}^{p} \rangle = \vert q_n^{c} \rangle \otimes \vert q_{n-1}^{p}\rangle \otimes \ldots \otimes \vert q_{0}^{p} \rangle,
\label{eq:dtqw_state_littlendian}
\end{equation}

\noindent which represents the state where the coin is in the state $\vert q_n^{c} \rangle$ and the walker is in the position state $\vert q_{n-1}^{p} \ldots q_{0}^{p} \rangle$.

\subsection{Quantum circuit design}
\label{sec:qc_design}
The efficiency of a quantum circuit implementing a DTQW relies on the efficient implementation of the single time-step operator \eqref{eq:dtqw_step}, so, ultimately, on that of the conditional shift operator \eqref{eq:S_comput_basis}. As shown in Eq. \eqref{eq:S_matrix}, the latter involves the circulant matrices $P_0$ and $P_1$ introduced in Eq. \eqref{eq:P0_P1_matrix} and circulant matrices are known to be diagonalized by the quantum Fourier transform (QFT) matrix (see Appendix \ref{app:diag_circ}). The QFT of the computational basis is defined as
\begin{equation}
{\rm QFT}: \;  \vert j \rangle  \; \mapsto \; \frac{1}{\sqrt{N}} \sum_{k=0}^{N-1}\omega_N^{ jk}\vert k \rangle,
\label{eq:QFT}
\end{equation}

\noindent where $\omega_N = \exp(2 \pi i/N)$ and $j,k=0,\ldots, N-1$, with matrix representation $\mathcal{F}_{j,k} = \omega_N^{jk}/\sqrt{N}$. The inverse QFT (IQFT) is represented by ${\mathcal{F}^\dagger}_{j,k} = \omega_N^{-jk}/\sqrt{N}$. The QFT is a unitary transformation, $\mathcal{F}\mathcal{F}^\dagger = \mathcal{F}^\dagger \mathcal{F}= I$. Accordingly, we can write

\begin{equation}
P_0 = \mathcal{F}^\dagger \Omega^\dagger \mathcal{F}, \quad\text{and} \quad P_1 = \mathcal{F}^\dagger \Omega \mathcal{F},
\label{eq:diag_P0_P1_qft}
\end{equation}

\noindent where
\begin{equation}
\Omega = \operatorname{diag}\left (1, \omega_N^{1},\ldots, \omega_N^{N-1} \right )
= R_1 \otimes R_2 \otimes \cdots \otimes R_n = \bigotimes_{k=1}^{n}R_k,
\label{eq:Omega}
\end{equation}

\noindent with 
\begin{equation}
R_k = \begin{pmatrix}
1 & 0\\
0 & e^{\frac{2 \pi i}{2^k}}
\end{pmatrix}
= \begin{pmatrix}
1 & 0\\
0 & \omega_N^{2^{n-k}}
\end{pmatrix}.
\label{eq:Rgate}
\end{equation}

\noindent We stress that $\Omega \vert j_p \rangle = R_1 \vert q_{n-1}^{p}\rangle \otimes \ldots  R_n \vert q_{0}^{p}\rangle$ (order matters) and that we can write the second equality of Eq. \eqref{eq:Omega} because we are assuming $N=2^n$.
Given Eq. \eqref{eq:diag_P0_P1_qft}, the conditional shift matrix $S$ \eqref{eq:S_matrix} and its diagonal form $\Sigma$ are related via
\begin{equation}
\Sigma = (I_c \otimes \mathcal{F}) S (I_c \otimes \mathcal{F}^\dagger)
= \vert 0_c \rangle\langle 0_c \vert \otimes \Omega^\dagger +\vert 1_c \rangle\langle 1_c \vert \otimes \Omega =
\begin{pmatrix}
\Omega^\dagger & \mathbf{0}\\
\mathbf{0} & \Omega
\end{pmatrix} 
\label{eq:Seig_QFT_S_IQFT},
\end{equation}

\noindent where the identity in coin space $I_c$ is required since the (I)QFT acts on position space only.

\begin{figure}[t]
\includegraphics[width=0.9\textwidth]{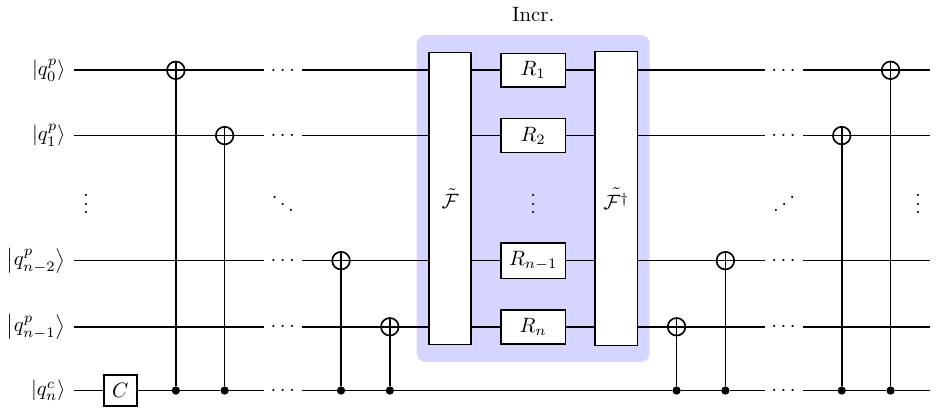}
\caption{Quantum circuit implementing one time-step of the DTQW on the $2^n$-cycle proposed in \cite{Shakeel2020}. The quantum Fourier transform, $\tilde{\mathcal{F}}$, and its inverse, $\tilde{\mathcal{F}}^\dagger$, do \textit{not} include the SWAP gates. The increment gate is diagonalized by the QFT [see also Fig. \ref{fig:ID_qc}(d)]. Conditional shift operator as in Fig. \ref{fig:ID_qc}.
\label{fig:QFT_qc}}
\end{figure}

The DTQW of $t$ steps is generated by repeatedly applying the operator $U$ $t$ times, Eq. \eqref{eq:psit}. Recalling that the QFT is unitary, it acts only on position space, and given that $S = (I_c \otimes \mathcal{F}^\dagger) \Sigma (I_c \otimes \mathcal{F})$ and that $(C \otimes I_p) = (I_c \otimes \mathcal{F}^\dagger) (C \otimes I_p) (I_c \otimes \mathcal{F})$, we can write
\begin{align}
U^t &= (I_c \otimes\mathcal{F}^\dagger) \left[ \Sigma ( C \otimes I_p)\right]^t (I_c \otimes \mathcal{F}),
\label{eq:dtqw_T_matrix}
\end{align}

\noindent with $\Sigma$ in Eq. \eqref{eq:Seig_QFT_S_IQFT}. Equation \eqref{eq:dtqw_T_matrix} provides a first sketch of the circuit we are going to implement. First, we perform a QFT on the position register, $I_c \otimes \mathcal{F}$. Then, we repeat $t$ times the single time-step evolution in the extended Fourier space (extended to include coin space), $\Sigma( C \otimes I_p)$. In the end, we perform an IQFT on the position register, $I_c \otimes \mathcal{F}^\dagger$. Even at this stage, an advantage of our scheme  is evident: Overall, it requires only one QFT and one IQFT, unlike the QFT-scheme in Fig. \ref{fig:QFT_qc}, which requires both the transformations at \textit{each} time step  \cite{Shakeel2020}.

Now, we focus on $\Sigma$ to further improve the above scheme. The second equality in Eq. \eqref{eq:Seig_QFT_S_IQFT} clearly shows the action of $\Sigma$: If the coin is in the state $\vert 0_c \rangle$ ($\vert 1_c \rangle$), then the operator $\Omega^\dagger$ ($\Omega$) acts on the position state. It is evident that, in the present form, each step of the DTQW requires $2n$ controlled-$R_k$ gates. To reduce the number of controlled operations, we point out that performing $R_k^\dagger = \operatorname{diag}(1,\exp\big[-2\pi i /2^k\big])$ on the $k$-th position qubit if the coin is in the state $\vert 0_c \rangle$ and $R_k = \operatorname{diag}(1,\exp\big[2\pi i /2^k\big])$ (opposite phase) if in $\vert 1_c \rangle$ is equivalent to performing $R_k^\dagger$ independently of the coin state, followed by a controlled-$R_k^2$ if the coin is in $\vert 1_c \rangle$ to compensate the previously assigned phase and get the correct one. Formally, we can rewrite the diagonal conditional shift operator \eqref{eq:Seig_QFT_S_IQFT} as
\begin{equation}
\Sigma = (\vert 0_c \rangle \langle 0_c \vert \otimes I_p + \vert 1_c \rangle \langle 1_c \vert \otimes \Omega^2) (I_c \otimes \Omega^\dagger),
\label{eq:S_proposed}
\end{equation}

\noindent where
\begin{equation}
\Omega^2 = \bigotimes_{k=1}^{n}R_k^2 = \bigotimes_{k=0}^{n-1}R_k = I \otimes R_1 \otimes \cdots \otimes R_{n-1},
\label{eq:Omega2}
\end{equation}

\noindent since $R_k^2  = R_{k-1}$ and $R_0 = I$ (one-qubit identity gate), see Eq. \eqref{eq:Rgate}.

\begin{figure}[t]
\includegraphics[width=0.8\textwidth]{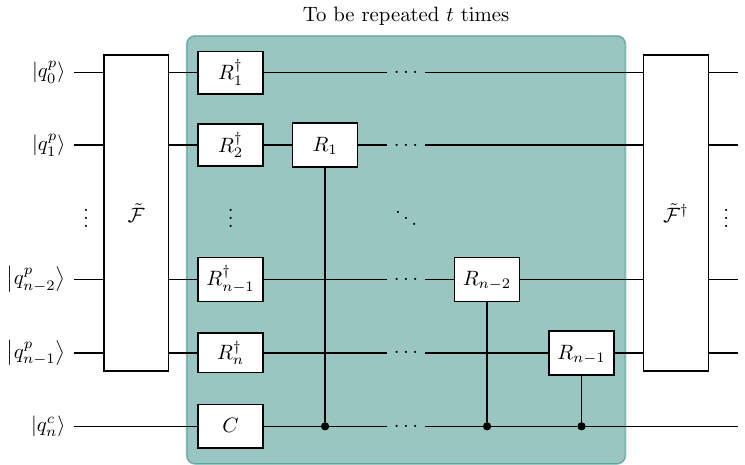}
\caption{Quantum circuit implementing one time-step of the DTQW on the $2^n$-cycle proposed in the present work. The quantum Fourier transform, $\tilde{\mathcal{F}}$, and its inverse, $\tilde{\mathcal{F}}^\dagger$, do \textit{not} include the SWAP gates. To implement $t$ time steps of the DTQW we have to concatenate the above circuit $t$ times. In doing so, since the QFT is unitary $\tilde{\mathcal{F}}^\dagger \tilde{\mathcal{F}} = I$, we are left with only one QFT at the beginning, the central block (shaded) repeated $t$ times, and one IQFT at the end. This simplification cannot occur in the quantum circuit in Fig. \ref{fig:QFT_qc} due to the CNOT and the coin gates. For an initially localized walker, $\vert \psi_0 \rangle = \vert \phi_c \rangle \otimes \vert 0_p \rangle$ the initial QFT is conveniently replaced by a layer of Hadamard gates (see Appendix \ref{app:opt_init_state}). 
\label{fig:our_qc}}
\end{figure}

As a final step in optimizing our quantum circuit, we want to make the SWAP operations, usually required in a proper (I)QFT to obtain the correct states, unnecessary. Therefore, following the argument in \cite{Shakeel2020}, it is useful to introduce the SWAP operation on the $n$-qubit register, which we denote by $\tau$. The SWAP takes qubit $k$ to qubit $n-1-k$ and \textit{vice versa},
\begin{equation}
\tau: \;  \vert q_{n-1} \ldots q_{0} \rangle \mapsto \vert q_{0} \ldots q_{n-1} \rangle.
\label{eq:swap_tau}
\end{equation}

\noindent This operation is unitary, $\tau^{-1} = \tau^\dagger$, and involutory, $\tau^{-1} = \tau$. A proper (I)QFT on $n$ qubits requires a SWAP on the $n$-qubit register at the end (beginning) of the circuit, i.e.,
\begin{align}
\mathcal{F} = \tau \tilde{ \mathcal{F} }, \quad \text{and} \quad \mathcal{F}^\dagger = \tilde{ \mathcal{F} }^\dagger \tau,
\label{eq:i_qft_tilde}
\end{align}

\noindent where $\tilde{ \mathcal{F} }^{(\dagger)}$ denotes the (I)QFT without the SWAP. Similarly, we introduce
\begin{equation}
\tilde{\Omega} = \tau \Omega \tau = R_{n} \otimes R_{n-1} \otimes \cdots \otimes R_{1},
\label{eq:Omega_tilde}
\end{equation}

\noindent see Eq. \eqref{eq:Omega}. Using Eqs. \eqref{eq:i_qft_tilde}-\eqref{eq:Omega_tilde} and recalling that $\tau^{-1}=\tau^\dagger=\tau$ acts only on the position register (not on the coin), we observe that
\begin{align*}
(I_c \otimes \mathcal{F}^\dagger) &= (I_c \otimes \tilde{\mathcal{F}}^\dagger)(I_c \otimes \tau),\\
(I_c \otimes \mathcal{F}) &= (I_c \otimes \tau)(I_c \otimes \tilde{\mathcal{F}}),\\
\Sigma & = (I_c \otimes \tau) (\vert 0_c \rangle \langle 0_c \vert \otimes I_p + \vert 1_c \rangle \langle 1_c \vert \otimes \tilde{\Omega}^2)(I_c \otimes \tilde{\Omega}^\dagger)(I_c \otimes \tau) ,\\
(C \otimes I_p) &= (I_c \otimes \tau) (C \otimes I_p) (I_c \otimes \tau),
\end{align*}

\noindent according to which we can rewrite Eq. \eqref{eq:dtqw_T_matrix} as 
\begin{align}
U^t &= (I_c \otimes \tilde{\mathcal{F}}^\dagger) \left[ (\vert 0_c \rangle \langle 0_c \vert \otimes I_p + \vert 1_c \rangle \langle 1_c \vert \otimes \tilde{\Omega}^2) (C \otimes \tilde{\Omega}^\dagger)\right]^t (I_c \otimes \tilde{\mathcal{F}}).
\label{eq:dtqw_T_matrix_swapless}
\end{align}

In conclusion, the quantum circuit implementing $t$ steps of a DTQW, excluding the initial state preparation, is shown in Fig. \ref{fig:our_qc} and does not need the SWAP in the (I)QFT. Size and depth of a quantum circuit implementing a DTQW can be further reduced by choosing a proper encoding of the position space and designing initial-state dependent circuits \cite{PhysRevA.104.062401}. We point out that the design of our quantum circuit is independent of the initial state, but it can be further optimized for an initially localized walker, the usual initial condition, by replacing the initial QFT with a layer of Hadamard gates (Appendix \ref{app:opt_init_state}).

\subsection{Comparison with other existing schemes}
\label{sec:qc_comparison}
In this section we estimate the size of the proposed quantum circuit (Fig. \ref{fig:our_qc}), in terms depth $\mathcal{D}$, number of one- and two-qubit gates, $\mathcal{N}^{(1)}$ and $\mathcal{N}^{(2)}$ respectively, and compare it with that of other existing schemes, following the preliminary analysis provided in \cite{Shakeel2020}. We will compare our scheme with the following ones. (i) The ID-scheme \cite{PhysRevA.79.052335}, which is based on the increment and decrement gates (Fig. \ref{fig:ID_qc}) that require generalized CNOT gates. The latter can be implemented in different ways, so, as an example, we consider their implementation (i.a) via linear-depth quantum circuit \cite{PhysRevA.87.062318} or (i.b) via ancilla qubits \cite{PhysRevA.52.3457}. In passing, we also mention a possible implementation via rotations \cite{PhysRevA.103.022408}. In the following analysis we consider the ID-scheme implemented as in Fig. \ref{fig:ID_qc}(d), with the increment gate only. (ii) The QFT-scheme \cite{Shakeel2020}, which is based on the increment gate diagonalized by the QFT (Fig. \ref{fig:QFT_qc}).

\begin{figure}[H]
\includegraphics[width=\textwidth]{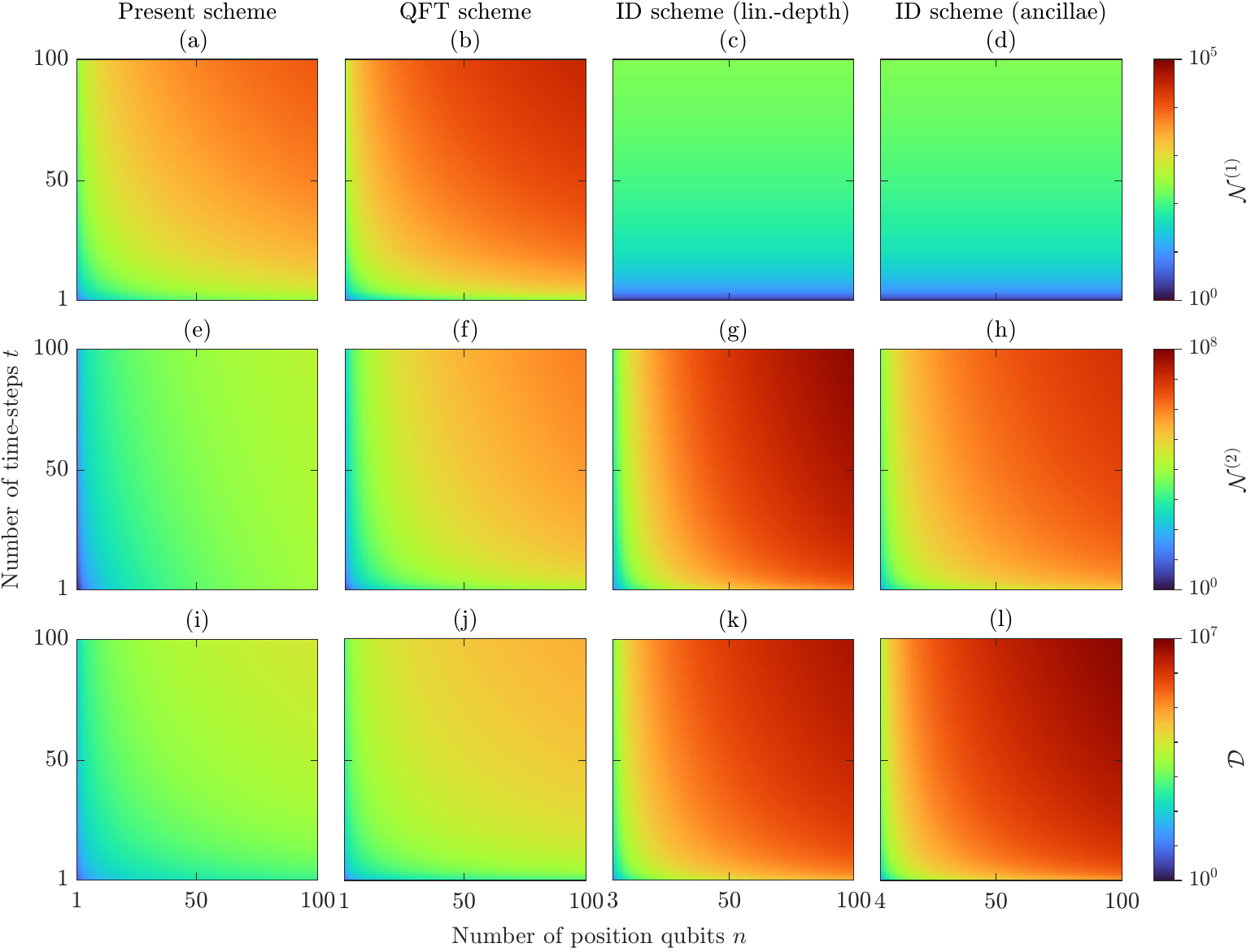}
\caption{Metrics in Table \ref{tab:qc_metrics} as a function of the number of position qubits, $n$, and the number of time-steps, $t$, of a DTQW on the $2^n$-cycle. Each column corresponds to a different scheme (Present, QFT, ID (lin.-depth), and ID (ancillae)) and each row to a different metric (number of one- $\mathcal{N}^{(1)}$ and two-qubit gates $\mathcal{N}^{(2)}$, circuit depth $\mathcal{D}$).
\label{fig:qc_metrics}}
\end{figure}

This discussion is neither supposed to be exhaustive, e.g., several are the ways to implement the generalized CNOT gates in the ID-scheme, nor to provide optimal and universal metrics, as the latter are ultimately quantum-device-dependent, e.g., it suffices to think of the process of transpilation, which rewrites and/or optimizes a given circuit according to the topology of the quantum device considered. Still, our estimate is instructive to assess how our scheme scales better than others when including the number of implemented time-steps in the analysis, in particular if compared to the QFT-scheme (Fig. \ref{fig:QFT_qc}) which is, to the best of our knowledge, the most efficient state-of-the-art implementation of DTQW on $2^n$-cycles. Results on the circuit size in the different schemes are summarized in Table \ref{tab:qc_metrics} and shown in Fig. \ref{fig:qc_metrics}. Details on the computation are deferred to Appendix \ref{app:analysis_dtqw_qc}.

Considering both the number $n$ of position qubits and the number $t$ of time-steps, the ID-scheme is the most resource-demanding among those examined: (i.a) If generalized CNOT gates are implemented via linear-depth quantum circuits, then circuit's depth increases as $\mathcal{D}= O(4 t n^2)$ and the results quickly degrade due to the large number of two-qubit gates, $\mathcal{N}^{(2)} = O(2 t n^3/3)$; (i.b) If generalized CNOT gates are implemented via ancilla qubits, then circuit's depth is of the same order, $\mathcal{D}= O(8 t n^2)$, but the number of two-qubit gates is reduced by an order, $\mathcal{N}^{(2)} = O(10 t n^2)$, at the cost of requiring extra qubits, $\mathcal{N}^{(a)} = O(n)$. Both the ID-approaches require $\mathcal{N}^{(1)} = 2t$. (ii) A remarkable improvement is obtained by the QFT-scheme, which, with no need of ancilla qubits, has $\mathcal{D}= O(6 t n)$ and $\mathcal{N}^{(2)} = O(t n^2)$, at the cost of increasing the number of one-qubit gates, $\mathcal{N}^{(1)} = O(3 t n)$. Our scheme refines such metrics by making the cost of the (I)QFT independent of the number of time-steps: In the long-time limit, $t \gg n$, we have $\mathcal{D}=\mathcal{N}^{(1)}=\mathcal{N}^{(2)} = O(tn)$, to which is addded a fixed, $t$-independent but $n$-dependent, cost $\mathcal{D} = O(4n)$,  $\mathcal{N}^{(2)} = O(n^2)$, and $\mathcal{N}^{(1)} = O(2n)$. To ease the comparison among the schemes, the metrics of Table \ref{tab:qc_metrics} are shown in Fig. \ref{fig:qc_metrics}, making it clear that our scheme outperforms the others in the number of two-qubit gates, which take the longest time to execute and are the noisiest in quantum computers. Also, we observe that the circuit depth is mainly determined by the number of two-qubit gates.

In conclusion, both the QFT-scheme and ours outperform the ID-scheme at any time. Although these metrics are comparable in the few time-steps regime, our scheme outperforms the QFT-scheme when a large number of time-steps is implemented.

\begin{table}[H]
\caption{Metrics of the quantum circuit implementing $t$ time-steps of a DTQW on the $2^n$-cycle for different schemes: $\mathcal{N}^{(1)}$ and $\mathcal{N}^{(2)}$ denote the number of one- and two-qubit gates,respectively, $\mathcal{D}$ the depth of the circuit, and $\mathcal{N}^{(a)}$ the number of ancilla qubits. The number $n$ refers to the number of qubits encoding walker's position. See also Fig. \ref{fig:qc_metrics}.
\label{tab:qc_metrics}}
\begin{adjustwidth}{-\extralength}{0cm}
\renewcommand{\arraystretch}{1.5}
\begin{tabularx}{\fulllength}{lccccc}
\toprule
\textbf{Scheme}
	& \textbf{Fig.}
	& \textbf{One-qubit} $\mathcal{N}^{(1)}$
	& \textbf{Two-qubit} $\mathcal{N}^{(2)}$
	& \textbf{Depth} $\mathcal{D}$ 
	& \textbf{Ancillae} $\mathcal{N}^{(a)}$\\
	\midrule
Present work
	& \ref{fig:our_qc}
	& $t(n+1) + 2n$
	& $t (n-1) + n(n-1)$
	& $ t n + 2(2n-1)$
	& n/a\\
QFT \cite{Shakeel2020}
	& \ref{fig:QFT_qc}
	& $t(3n+1)$
	& $t n(n+1)$
	& $t (6n)$
	& n/a\\
ID* \cite{PhysRevA.79.052335} (lin.-depth q.c. \cite{PhysRevA.87.062318}, $n \geq 3$)
	& \ref{fig:ID_qc}(d)
	& $2t$
	& $t\frac{1}{3}(2n^3-6n^2+13n-3)$
	& $t(4n^2-14n+19 )$
	& n/a\\
ID* \cite{PhysRevA.79.052335} (ancillae  \cite{PhysRevA.52.3457}, $n \geq 4$)
	& \ref{fig:ID_qc}(d)
	& $2t$
	& $t (10n^2-48n+66) $
	& $t(8n^2-38n+55)$
	& $n-3$\\
\bottomrule
\end{tabularx}
\end{adjustwidth}
\noindent{\footnotesize{* These schemes differ in how the generalized CNOT gates in the increment gate are realized [see Fig. \ref{fig:ID_qc}(b)].}}
\end{table}

\section{Results and Discussion}
\label{sec:results}
We test the DTQW circuit introduced in Sec. \ref{sec:qc_design} (see Fig. \ref{fig:our_qc}) on the IBM quantum computer \texttt{ibm\_cairo} v.1.3.5, a 27-qubit Falcon r5.11 processor whose qubit connectivity map is shown in Fig. \ref{fig:ibm_cairo}(a). In the following, we introduce the DTQW considered for the test and the quantities of interest, then we point out some solutions to improve the circuit, and finally we present and discuss the results.

\subsection{Hadamard DTQW}
\label{sec:H-DTQW}
A common choice for the coin operator is the Hadamard coin
\begin{equation}
C = \frac{1}{\sqrt{2}}
\begin{pmatrix}
1 & 1\\
1 & -1
\end{pmatrix}.
\label{eq:Hadamard_coin}
\end{equation}

\noindent As for the initial state, we assume the walker to be initially localized in $\vert 0_p\rangle$ and the coin to be in a given superposition
\begin{equation}
\vert \psi_0 \rangle = \left[ \cos \left(\frac{\pi}{12} \right) \vert 0_c \rangle + i \sin \left(\frac{\pi}{12} \right) \vert 1_c \rangle \right] \vert 0_p \rangle.
\label{eq:psi0_ibm}
\end{equation}

\noindent The Hadamard DTQW for this initial state has the following properties: (i) The dynamics is periodic of period $T_{\rm dyn}= 8$ ($T_{\rm dyn}= 24$) on the 4-cycle (8-cycle) \cite{DUKES2014189}; (ii) Maximally entangled single-particle states---entanglement between position and coin---are generated after one step of the walk and then recurrently with period $T_{\rm MESPS} = 4$ ($T_{\rm MESPS}= 12$) on the 4-cycle (8-cycle) \cite{PhysRevA.108.L020401}. These properties are therefore suitable for thoroughly testing the quality of the designed quantum circuit, i.e., to assess to what extent the actual implementation can reproduce these ideal features.  We point out that, for a Hadamard DTQW on the 4- and 8-cycle, any initial state \eqref{eq:psi0} with $\phi = \pi / 2$ will generate a dynamics with the two above mentioned features for any value of $\theta$ \cite{PhysRevA.108.L020401}. We arbitrarily set $\theta = \pi/6$ to initialize the coin in a non-trivial superposition of states, see Eq. \eqref{eq:psi0_ibm}. For later convenience, we anticipate that the periodic dynamics of the DTQW in the 4- and 8-cycle sustain the periodic occurrence of localized states when the initial state is localized in position space. However, dynamics and occurrence of localized states can have different periods. In the 4-cycle, the initial state localized in $\vert 0 \rangle_p$ is perfectly transferred to $\vert 2 \rangle_p$ after $4$ time steps, hence localized states occur with period $T_{\rm dyn}/2=4$, in contrast with the period $T_{\rm dyn}=8$ of the dynamics. Instead, in the 8-cycle localized states occur only as a result of the periodic dynamics ($T_{\rm dyn}=24$) \cite{PhysRevA.108.L020401}.
Implementing the cycle with $N=4$ and $N=8$ vertices requires $n=2$ and $n=3$ position qubits, respectively.

\subsection{Figures of merit}
\textit{Probability distribution.---}In most DTQW problems, we are interested in the probability distribution of the walker's position. Our purpose is to compare the ideal probability distribution with the experimental ones, the latter obtained in a noisy simulation and in an actual implementation on the quantum hardware \texttt{ibm\_cairo}. To compare two discrete probability distributions, $P=\{p_k\}_k$ and $Q=\{q_k\}_k$, we adopt the Hellinger fidelity $\mathcal{H}(P,Q) = [1-h^2(P,Q)]^2$, where the Hellinger distance $h(P,Q)$ \cite{trevisan2023lecture} between $P$ and $Q$ is defined by
\begin{equation}
h(P,Q) = \frac{1}{\sqrt{2}} \sqrt{ \sum_{k} \left ( \sqrt{p_k} - \sqrt{q_k}\right )^2 }.
\label{eq:H2_dist}
\end{equation}

\noindent The Hellinger distance is symmetric, $h(P,Q)=h(Q,P)$, and bounded $0 \leq h(P,Q) \leq 1$, with $h=0$ meaning that the two distributions are equal (fidelity $\mathcal{H}=1$).

\textit{Entanglement.---}Usually, in a DTQW entanglement between walker and coin occurs. This can be understood as hybrid entanglement, also referred to as single-particle entanglement because  established between different degrees of freedom of the same quantum system, here position and coin, which is an internal degree of freedom, e.g., spin \cite{azzini2020spe}. Bipartite entanglement can be probed by means of entanglement entropies. In this case, we probe the second-order R\'{e}nyi entanglement entropy \cite{RevModPhys.81.865} of the reduced density matrix for the two parts (coin and position) via randomized measurements \cite{Brydges2019}. Estimating this quantity requires significantly fewer measurements than performing quantum state tomography: For a $n$-qubit state, $O(2^{a n})$ with $a<2$ (the coefficient $a$ depends on the nature of the considered state, see Supplementary Materials for \cite{Brydges2019}) compared to $2^{2n}-1$ \cite{PhysRevA.66.012303}. In this regard, the advantage of this approach becomes remarkable when a large number of qubits is involved---as expected to be in future applications---due to the current costliness of tomography. The second-order R\'{e}nyi entropy for a part $A$ of the total bipartite system described by $\rho_{AB}$ is defined as
\begin{equation}
    S^{(2)}(\rho_A) = -\log_2 \Tr\big[\rho_A^2\big],
\end{equation}

\noindent with $\rho_A = \Tr_B \rho_{AB}$ the reduced density matrix for part $A$. If the second-order R\'{e}nyi entropy of the part is greater than that of total system, $S^{(2)}(\rho_A) > S^{(2)}(\rho_{AB})$, then bipartite entanglement exists between the two parts (for separable states $S^{(2)}(\rho_A) \leq S^{(2)}(\rho_{AB})$ and $S^{(2)}(\rho_B) \leq S^{(2)}(\rho_{AB})$ \cite{RevModPhys.81.865}). If, in addition, the overall state $\rho_{AB}$ is pure, then the second-order R\'{e}nyi entropy is directly a measure of bipartite entanglement and $S^{(2)}(\rho_{A})=S^{(2)}(\rho_{B})$ (the reduced density matrices of a pure bipartite state have the same non-zero eigenvalues, from the Schmidt decomposition). The second-order R\'{e}nyi entropy is maximum for the maximally mixed state, $\max_{\rho_A} S^{(2)}(\rho_A) = \log_2 d_A$, with $d_A$ the dimension of $\rho_A$. Furthermore, $S^{(2)}(\rho_{AB})$ is indicative of the overall purity of the system, because it is null for pure quantum states.

\subsection{Circuit optimization}
\begin{figure}[t]
\includegraphics[width=\textwidth]{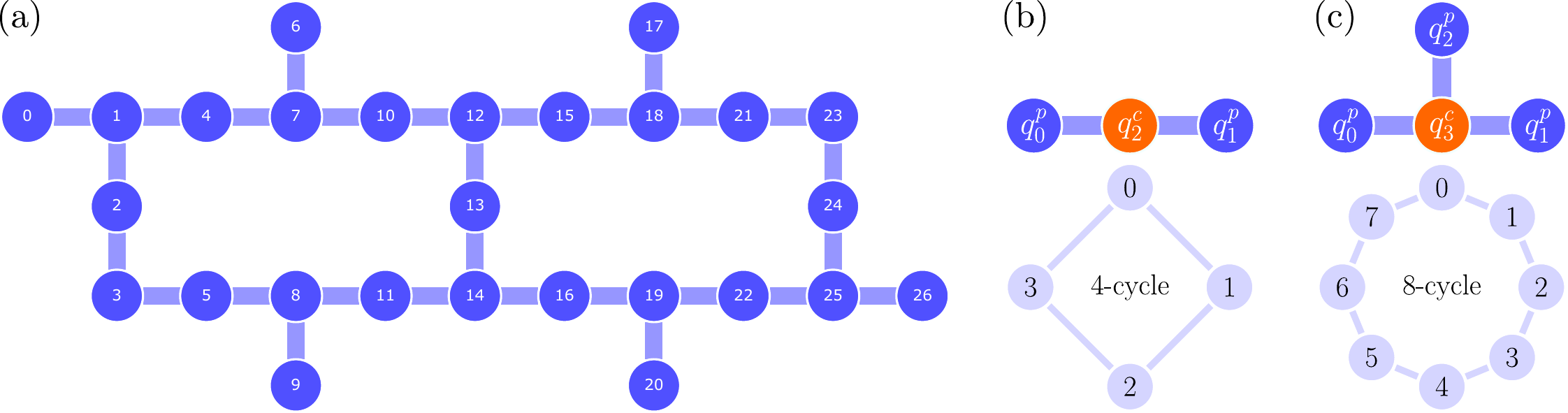}
\caption{\textbf{(a)} Qubit connectivity map of \texttt{ibm\_cairo}. \textbf{(b,c)} Optimal mapping of the coin-position state onto multiqubit state $\vert q_n^{c} q_{n-1}^{p}\ldots q_{0}^p \rangle$ for the cycle with (b) $N=4$ and (c) $N=8$ vertices ($n=2,3$ position qubits, respectively). No SWAP operations between position and coin qubits are required by the controlled-$R_k$ gates in Fig. \ref{fig:our_qc}, the coin qubit (orange) being already adjacent to all position qubits (blue).
\label{fig:ibm_cairo}}
\end{figure}

Each time step of the DTQW requires the coin qubit to interact with each position qubit (see controlled-$R_k$ gates in Fig. \ref{fig:our_qc}). The  quantum hardware may have limited connectivity and whenever two qubits are not physically adjacent, then SWAP operations are needed to make them interact. A wise circuital implementation must account for the  connectivity of the quantum hardware considered. We can limit the number of SWAPs by making the coin qubit as ``shared'' as possible, compatibly with the typical sparse connectivity of superconducting quantum computers. The qubit topology of \texttt{ibm\_cairo} [Fig. \ref{fig:ibm_cairo}(a)] makes it possible to implement the circuit for a DTQW on the 4- and 8-cycle without SWAPs by mapping coin and position qubits as in Fig. \ref{fig:ibm_cairo}(b) and (c), respectively. Given the optimal mapping compatible with the given qubit connectivity, we consider the set of qubits having the lowest error rates averaged over different calibrations.
In addition, the initial state \eqref{eq:psi0_ibm} is localized in position space, therefore we replace the initial QFT with a layer of Hadamard gates (Appendix \ref{app:opt_init_state}).

\subsection{Analysis of the DTQW on the 4- and 8-cycle}

\begin{figure}[t]
\includegraphics[width=1\textwidth]{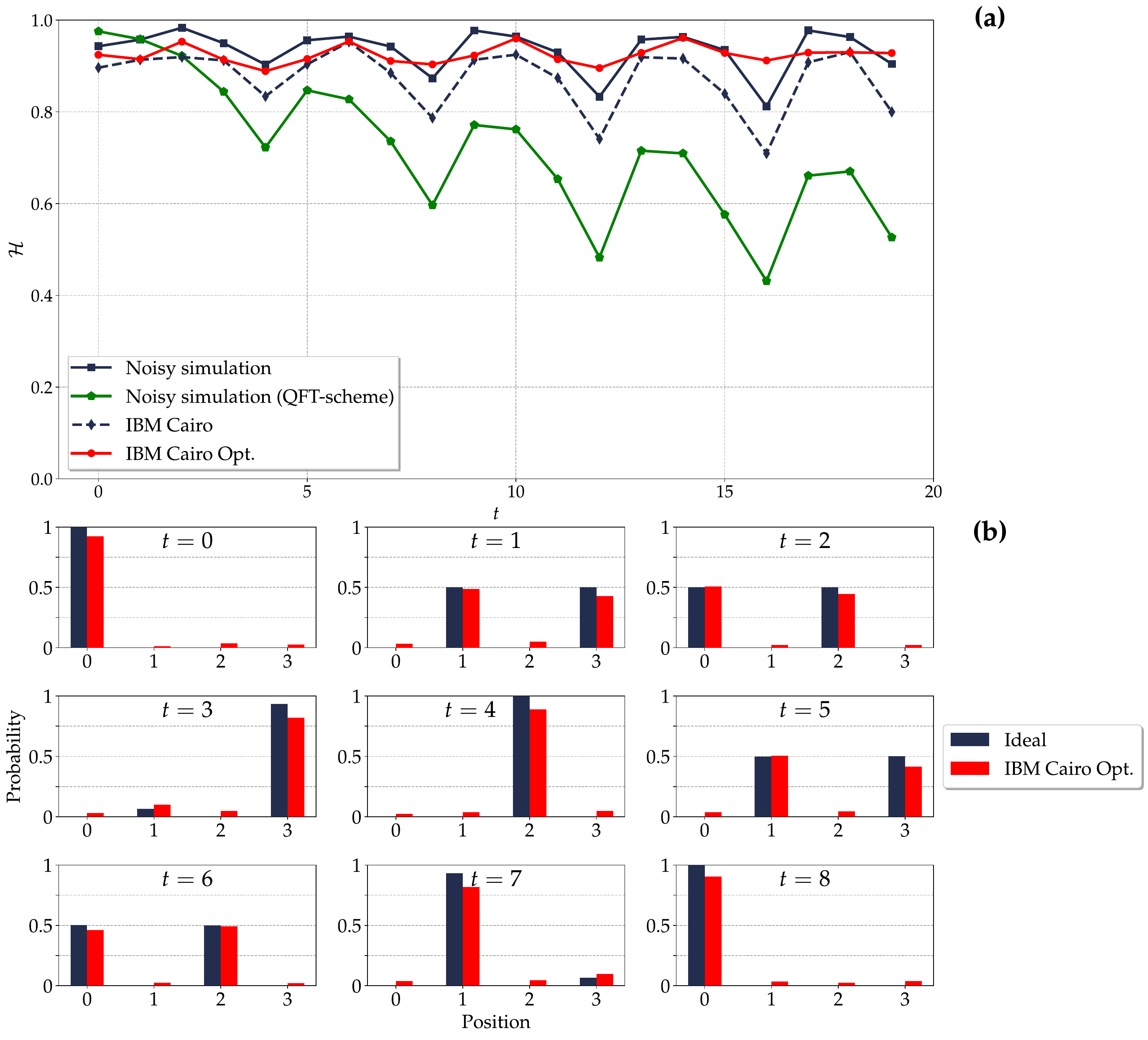}
\caption{Hadamard DTQW on the $(N=4)$-cycle with initial state as in Eq. \eqref{eq:psi0_ibm}. \textbf{(a)} Hellinger fidelity between the ideal and the experimental probability distributions of walker's position as a function of time-step $t$. The experimental distributions include: Noisy simulation and implementations on the actual quantum hardware with \texttt{optimization\_level=1} (IBM Cairo) and \texttt{optimization\_level=3} (IBM Cairo Opt.) in transpilation. Results for the noisy simulation of the DTQW circuit in the QFT-scheme \cite{Shakeel2020} are reported for comparison. \textbf{(b)} Ideal and experimental (IBM Cairo Opt.) probability distributions of walker's position for time-steps $t=0,\ldots,8$.
Results for both simulations and quantum hardware implementation are obtained for $10^5$ shots and by encoding the position state in qubits number 3 and 8 and the coin state in qubit 5  of \texttt{ibm\_cairo}, see Fig. \ref{fig:ibm_cairo}(a),(b).
\label{fig:n2_HellingerFid}}
\end{figure}

\textit{Probability distribution (4-cycle).---}Performing a noisy simulation of the quantum circuit is the preliminary step before the actual implementation on the quantum hardware. Fig. \ref{fig:n2_HellingerFid}(a) shows the Hellinger fidelities of the walker's position distribution in the 4-cycle. The Hellinger fidelity for the noisy simulation of our circuit is above the $80\%$ for all the 19 time steps implemented, while the results for the noisy simulation of the circuit in the QFT-scheme \cite{Shakeel2020} degrade below the $80\%$ after a few steps. The previous analysis on the circuit size proved the advantage of our circuit in the long time limit, but these results suggest that our circuit outperforms the QFT-scheme circuit already in the few time-steps regime. For the actual implementation on \texttt{ibm\_cairo} we consider two levels of optimization in the transpilation:
(i) \texttt{optimization\_level=1} (default value), which transpiles the circuit into the native gates of the hardware and performs a light optimization (blue dashed line, curve ``IBM Cairo''), and (ii) \texttt{optimization\_level=3}, which performs the heaviest optimization (red solid line, curve ``IBM Cairo Opt.'').

The Hellinger fidelity for the default implementation of our circuit (\texttt{op\-ti\-miza\-tion\-\_\-lev\-el=1}) shows a moderate discrepancy with respect to the noisy simulation, but closely follows the trend of the latter. The heavily optimized implementation of our circuit (\texttt{optimization\_level=3}) provides better results and also partially mitigates the local minima of the noisy simulation. However, we point out that the circuit transpiled with the highest optimization level turns out to have a depth which is basically independent of the number of time-steps implemented (see Appendix \ref{app:transpiled_dtqw_qc}). This explains the long-lasting optimality of the results, $\mathcal{H} \gtrsim 90\%$ up to $t=19$ (last time-step implemented). However, this optimal transpiled circuit is obtained only for $n=2$ position qubits (4-cycle);  for $n=3$ (8-cycle) we obtain a circuit whose depth increases with the number of time-steps. The Hellinger fidelity at $t=0$ is lower than 1 because we still implement the whole circuit with $t=0$, i.e., we do not just implement the initial state. Also, the Hellinger fidelity is characterized by periodic local minima, with period 4 which is half of the period of the dynamics. These minima occur when the walker is ideally localized in position, a probability distribution so peaked (delta) that it can hardly be obtained as the result of an actual, noisy implementation [see Fig. \ref{fig:n2_HellingerFid}(b)]. The frame  $t=8$ in panel (b) shows that the walker has returned to the initial position, as expected for the periodic dynamics.

\begin{figure}[t]
\includegraphics[width=1\textwidth]{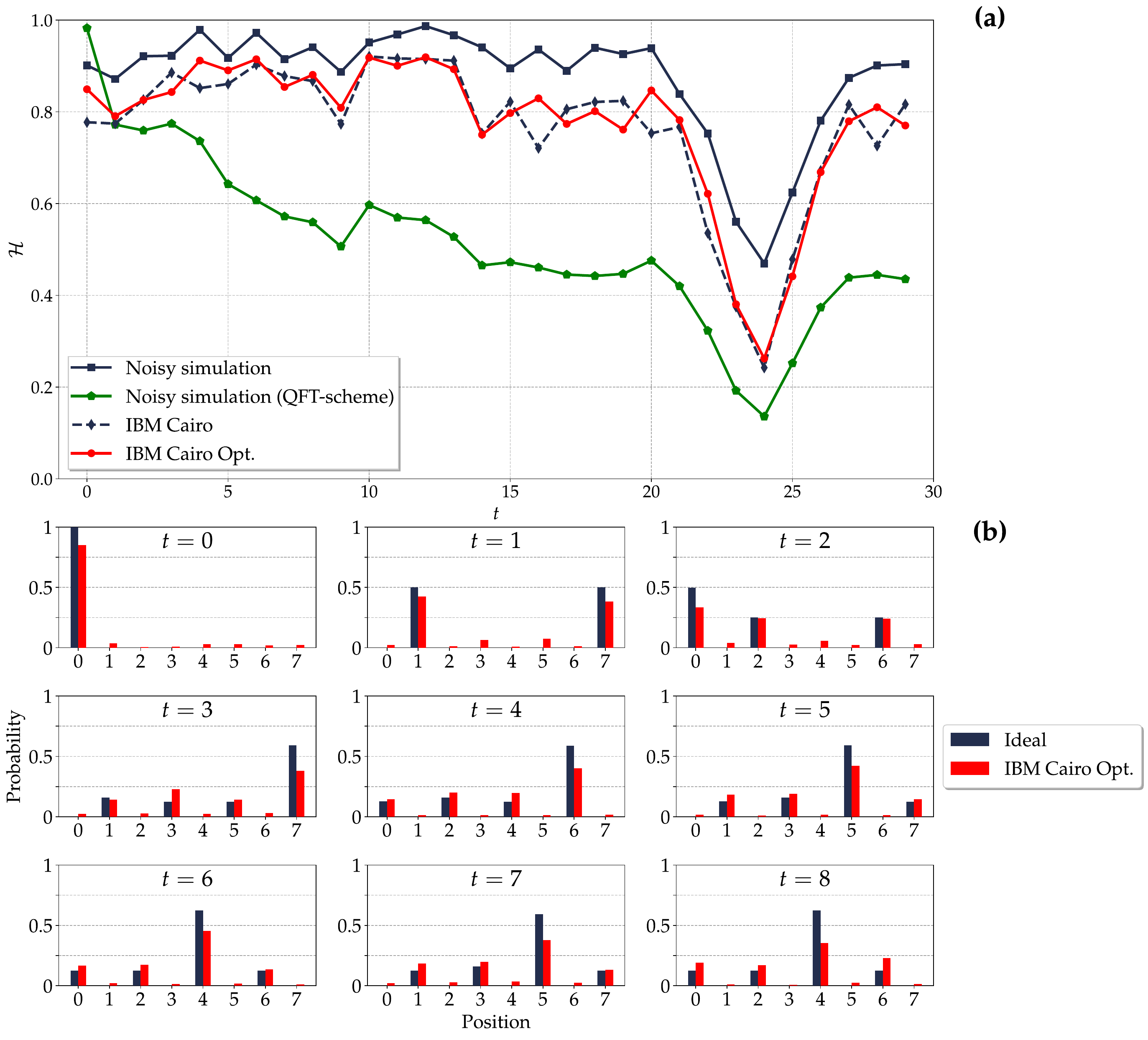}
\caption{Same as in Fig. \ref{fig:n2_HellingerFid} but for $N=8$. Results for both simulations and quantum hardware implementation are obtained for $10^5$ shots and by encoding the position state in qubits number 10, 13, and 15 and the coin state in qubit 12 of \texttt{ibm\_cairo}, see Fig. \ref{fig:ibm_cairo}(a),(c).
\label{fig:n3_HellingerFid}}
\end{figure}

\textit{Probability distribution (8-cycle).---}We obtain qualitatively analogous results also for the DTQW on the 8-cycle (Fig. \ref{fig:n3_HellingerFid}). However, as anticipated in the previous paragraph, in this case the optimization in the transpilation is not as effective as for the 4-cycle. The Hellinger fidelities with \texttt{optimization\_level=1} and \texttt{3} are thus consistent with each other. Unlike the DTQW on the 4-cycle, given an initial localized state, localized states at later times occur only as a result of the periodic dynamics ($T_{\rm dyn}=24$), hence the minimum of the Hellinger fidelity at $t=24$.

\begin{figure}[t]
\includegraphics[width=0.85\textwidth]{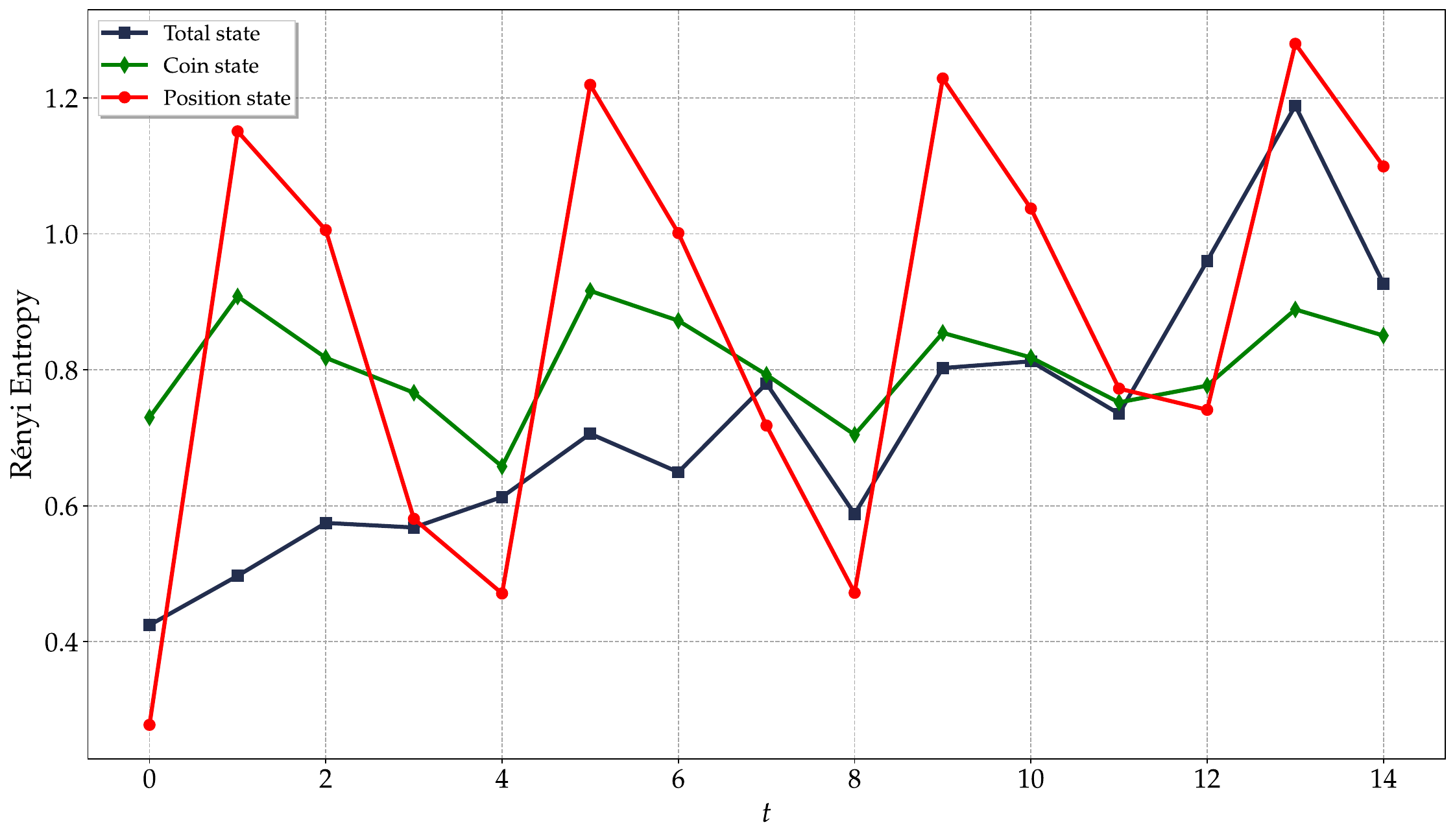}
\caption{Recurrent generation of maximally entangled single-particle states for a Hadamard DTQW on the $(N=4)$-cycle with initial state as in Eq. \eqref{eq:psi0_ibm} investigated by means of the second-order R\'{e}nyi entropy as a function of time-steps $t$. Bipartite entanglement between the two parts (coin and position degrees of freedom) exists if the second-order R\'{e}nyi entropy of a part is larger than that of the total system. Results of entropies are obtained via 300 randomized measurements \cite{Brydges2019} and $10^5$ shots for each step of the DTQW, with \texttt{optimization\_level=1} in transpilation. Position state is encoded in qubits number 3 and 8 and the coin state in qubit 5  of \texttt{ibm\_cairo}, see Fig. \ref{fig:ibm_cairo}(a),(b).
\label{fig:n2_Entropies}}
\end{figure}

\textit{Entanglement (4-cycle).---}Probing the second-order R\'{e}nyi entanglement entropy via randomized measurements \cite{Brydges2019} for many time-steps of the DTQW and for increasing size of the cycle is computationally demanding. Therefore, to discuss the recurrent generation of maximally entangled single-particle states, we focus only on the DTQW on the 4-cycle (Fig. \ref{fig:n2_Entropies}). First, we review the ideal scenario. Given an initial pure state and a unitary evolution, the 
second-order R\'{e}nyi entropy of coin state, $\rho_c$, and position state, $\rho_p$, are equal, $S^{(2)}(\rho_c)=S^{(2)}(\rho_p)$, and that of total system, $\rho_{cp}$, is null (pure state) at any time $t$. Also, we have $S^{(2)}_{\rm max}(\rho_c) =  S^{(2)}_{\rm max}(\rho_p) = \log_2(\min\{d_c,d_p\}) = 1$, the two-dimensional coin being the part with the lowest dimension. The initial state $\vert \psi_0\rangle$ in Eq. \eqref{eq:psi0_ibm} is pure and separable (no entanglement), as well as the state at $t=4$
\begin{equation}
\vert \psi_4 \rangle = \left[ \cos \left(\frac{\pi}{12} \right) \vert 0_c \rangle + i \sin \left(\frac{\pi}{12} \right) \vert 1_c \rangle \right] \vert 2_p \rangle,
\end{equation}

\noindent which is $\vert \psi_0\rangle$ perfectly transferred in position space from $\vert 0_p\rangle$ to $\vert 2_p\rangle$. The dynamics has period $T_{\rm dyn}=8$, thus $S^{(2)}(\rho_\alpha) = 0$ at $t = 4k$ with $k=0,1,2,\ldots$ and $\alpha = c,p$, denoting the absence of entanglement when periodic separable states occur. In addition, maximally entangled single-particle states are generated at $t=1$ and $t=5$
\begin{align}
    \vert \psi_1 \rangle &= \frac{1}{\sqrt{2}} \left[ e^{i \pi/12} \vert 0_c\rangle\vert 3_p\rangle + e^{-i \pi/12} \vert 1_c\rangle\vert 1_p\rangle \right],\\
    \vert \psi_5 \rangle &= \frac{1}{\sqrt{2}} \left[e^{i \pi/12} \vert 0_c\rangle\vert 1_p\rangle + e^{-i \pi/12} \vert 1_c\rangle\vert 3_p\rangle\right],
\end{align}

\noindent and are later supported by the periodic dynamics, thus $S^{(2)}(\rho_\alpha) = 1$ at $t = 1 + 4k$ with $k=0,1,2,\ldots$ and $\alpha = c,p$. For the remaining time-steps, $t = 2+4k$ and $t=3+4k$, partially entangled states are expected (see Supplementary Material of \cite{PhysRevA.108.L020401}).

Experimentally (Fig. \ref{fig:n2_Entropies}), the second-order R\'{e}nyi entropy of the total bipartite system is nonzero even at $t=0$ and increases with time, denoting a further degradation of the purity of the state. Since the latter is not pure, in general we expect $0<S^{(2)}(\rho_{cp}) \leq 3$ (the upper bound refers to the maximally mixed 3-qubit state), $S^{(2)}(\rho_c) \neq S^{(2)}(\rho_p)$ and different maximum values, $S^{(2)}(\rho_c) \leq \log_2 d_c = 1$ and $S^{(2)}(\rho_p) \leq \log_2 d_p = 2$ (4-cycle).
Whenever separable states are expected ($t=4k$), we observe local minima of $S^{(2)}(\rho_\alpha)$, but either part has $S^{(2)}(\rho_\alpha)>S^{(2)}(\rho_{cp})$, thus denoting the presence of residual entanglement. Whenever maximally entangled single-particle states are expected ($t=1+4k$), we observe local maxima of $S^{(2)}(\rho_\alpha)$, with $\alpha = c,p$, both of which are greater than $S^{(2)}(\rho_{cp})$ up to $t=9$. For the remaining time-steps, we observe that at least either part has $S^{(2)}(\rho_\alpha) \gtrsim S^{(2)}(\rho_{cp})$, consistently with the expected presence of partial entanglement.

To summarize, the actual multiqubit state implemented on the quantum hardware and then processed by the quantum circuit is generally mixed and characterized by residual entanglement, but the local minima and maxima of $S^{(2)}(\rho_c)$ and $S^{(2)}(\rho_p)$ are perfectly consistent with the expected periodic separable and maximally entangled single-particle states, respectively, up to $t=9$. Results suggest that the observed local maxima will preserve the expected periodicity for further steps $t>9$, but eventually $S^{(2)}(\rho_{cp}) \geq S^{(2)}(\rho_\alpha)$ due to the degradation of the purity of the total state.
Entanglement is a distinct quantum signature and, in this example, we have clear evidence of generation up to $t=9$. This does not necessarily imply that quantum features, including entanglement, in the realization of the DTQW are lost thereafter. Indeed, for the remaining time-steps investigated, $9 < t\leq 14$, we still observe $S^{(2)}(\rho_p) \gtrsim S^{(2)}(\rho_{cp})$---presence of entanglement---whenever states with partial or maximum entanglement are expected. Therefore, quantum features are present up to the last time-step considered, $t=14$.

\section{Conclusions}
\label{sec:conclusions}

We proposed an efficient quantum circuit for the DTQW on the $2^n$-cycle. Our scheme, using only one QFT and one IQFT, significantly improves the most efficient state-of-the-art implementation \cite{Shakeel2020} (QFT-scheme in the following), which uses one QFT and one IQFT at \textit{each} time step. As a result, our circuit requires only $O(n^2 + nt)$ two-qubit gates, compared to the $O(n^2 t)$ of the QFT-scheme. The improvement in this gate count is even more significant at long times, passing, for $t \gg n$, from $O(n^2 t)$ in the QFT-scheme to $O(n t)$ in ours. Therefore, two-qubit gates taking the longest time to execute and being the noisiest, our quantum circuit is computationally less demanding and also paves the way for reliable use on noisy-intermediate scale quantum devices \cite{preskill2018quantum,corcoles_2020,RevModPhys.94.015004}.

In this regard, we tested the proposed quantum circuit on an actual quantum hardware, \texttt{ibm\_cairo}, considering a Hadamard DTQW on the $4$- and $8$-cycle. Both are characterized by periodic dynamics \cite{DUKES2014189} and recurrent generation of maximally entangled single-particle states \cite{PhysRevA.108.L020401}. We claim two main results. First, even in the short-time regime, the present quantum circuit outperforms the current state-of-the-art DTQW circuits, whose results are degraded after only a few steps \cite{PhysRevA.103.022408,Shakeel2020}. Despite the moderate discrepancy, our implementation on an actual quantum hardware provides results that closely follow those from the noisy simulation. In particular, the Hellinger fidelity between the ideal probability distribution of walker's position and the experimental one is above $90\%$ for all the $t=19$ time-steps we implemented in the 4-cycle and above $80\%$ up to $t=13$ time-steps in the 8-cycle. Second, for the DTQW on the 4-cycle, we provide experimental evidence of the recurrent generation of nearly maximally entangled single-particle states up to $t=9$ time-steps. The expected maximum entanglement is not achieved because the ideally pure state of the bipartite system is actually implemented on the quantum hardware as a mixed multiqubit state and its purity degrades over time.
 
The implementation of our circuit on actual quantum computers may benefit from the following. The circuit strongly relies on controlled $R_k$-gates (phase shift gate) and the latter can be efficiently implemented using a single ancillary qubit \cite{Kim2018}. Moreover, as the position space increases, the sparse connectivity of a superconducting quantum computer results in large experimental overheads of SWAP gates, which becomes unavoidable, e.g., to make the coin qubit interact with each position qubit. In this regard, a virtual two-qubit gate can be employed to suppress errors due to the additional SWAP gates \cite{yamamoto2023}. Alternatively, an implementation on a quantum hardware architecture with full connectivity \cite{linke2017,murali2020,PRXQuantum.3.010344} may be more advantageous.

Possible applications of our scheme include the circuital implementation of direct communication protocols \cite{Srikara2020,PhysRevA.107.022611} and quantum key distribution protocols \cite{vlachou2018} based on DTQW on the cycle. We point out that the proposed circuit does not impose constraints on the coin operator (one-qubit gate), which in principle can be changed at each time step. Parrondo's paradox arises when losing strategies are combined to obtain a winning one and it cuts across various research areas. This counterintuitive phenomenon can be observed in DTQW on the line or cycle when two or more coin operators are applied in a deterministic sequence \cite{PhysRevE.104.064209,PhysRevA.104.012204}.
Therefore, we expect that our quantum circuit may also be of interest to quantum game theory  \cite{Lai2020}.

Circuit implementations of DTQW on a cycle of arbitrary $N$ have been addressed in \cite{Shakeel2020,Wing-Bocanegra2023}, while our proposal is limited to DTQW on the $N$-cycle with $N=2^n$. We point out, however, that any circulant matrix is diagonalized by the QFT and this has been already exploited to efficiently implement CTQWs on circulant graphs \cite{Qiang2016}. Similarly, our implementation for DTQWs can be generalized to circulant graphs---graphs whose adjacency matrix is circulant---which, being $d$-regular (each vertex has degree $d$), will require a $d$-dimensional coin \cite{Aharonov2001qw,Tregenna_2003}. A generalization of our approach to more complex structures is therefore desirable and potentially of larger interest, e.g., for algorithmic purposes.

\vspace{6pt} 


\authorcontributions{L.R. conceived the idea and supervised the work, with inputs from G.C, M.B. and G.B., and drafted the manuscript. G.C. performed quantum simulations by coding actual IBM quantum processors.  All authors discussed the results and contributed to revision of the manuscript. All authors have read and agreed to the published version of the manuscript.}

\funding{L.R., G.C., and G.B. acknowledge the financial support of the INFN through the project `QUANTUM'. M.B. acknowledges financial support from NQSTI within PNRR MUR project PE0000023-NQSTI.}

\institutionalreview{Not applicable.}

\informedconsent{Not applicable.}

\dataavailability{The dataset generated and analyzed in the current study is available from the corresponding author upon reasonable request.} 

\acknowledgments{We acknowledge the use of IBM Quantum services and the access to them granted by INFN for this work. The views expressed are those of the authors, and do not reflect the official policy or position of IBM or the IBM Quantum team.}

\conflictsofinterest{The authors declare no conflict of interest. The funders had no role in the design of the study; in the collection, analyses, or interpretation of data; in the writing of the manuscript; or in the decision to publish the results.} 

\abbreviations{Abbreviations}{
The following abbreviations are used in this manuscript:\\

\noindent 
\begin{tabular}{@{}ll}
CTQW	& Continuous-time quantum walk\\
DTQW	& Discrete-time quantum walk\\
(I)QFT	& (Inverse) quantum Fourier transform
\end{tabular}
}

\appendixtitles{yes} 
\appendixstart
\appendix

\section[\appendixname~\thesection]{Eigenvalues and eigenvectors of circulant matrices}
\label{app:diag_circ}
Let $C$ be a $N \times N$ circulant matrix \cite{gray2006toeplitz}
\begin{equation}
C =
\begin{pmatrix}
c_0		& c_1	& \ldots & c_{N-2}	& c_{N-1} \\
c_{N-1} & c_0		& c_1 & & c_{N-2}  \\
\vdots	& c_{N-1} & c_0 & \ddots & \vdots \\
c_2		&	& \ddots & \ddots	 & c_1 \\
c_1 & c_2  &\ldots & c_{N-1} & c_0
\end{pmatrix},
\label{eq:circ_matr}
\end{equation}

\noindent where each row is a cyclic shift of the row above it. The structure can also be characterized by noting that
\begin{equation}
C_{k,j} = c_{(j-k) \bmod N}.
\end{equation}

\noindent The eigenvalues $\lambda_m$ and eigenvectors $y^{(m)}$ (column vector) of $C$, solutions of $C y^{(m)} = \lambda_m y^{(m)}$, are
\begin{align}
\lambda_m &= \sum_{k=0}^{N-1} c_k \omega_N^{-mk}, \label{eq:eigval_circ}\\
y^{(m)} &= \frac{1}{\sqrt{N}}\left (1,\omega_N^{-1},\omega_N^{-2},\ldots,\omega_N^{-(N-1)}\right )^\intercal,
\end{align}

\noindent with $\omega_N = \exp(2\pi i/N)$ and the symbol ${}^\intercal$ denoting the transpose without complex conjugation. Letting $\Lambda = \operatorname{diag}(\{\lambda_m\}_m)$, we have
\begin{equation}
C = U \Lambda U^\dagger, \quad \text{i.e.,} \quad
\Lambda = U^\dagger C U,
\end{equation}

\noindent where $U$ is the unitary matrix
\begin{equation}
U = \left( y^{(0)} \left\vert y^{(1)} \right\vert \cdots \left\vert y^{(N-1)} \right.\right)
\end{equation}

\noindent whose elements read $U_{j,k} = \omega_N^{-j k}/\sqrt{N}$, with $j,k=0,1, \ldots,N-1$. Since they have the same definition, in $U^\dagger$ and $U$ we recognize, respectively, the quantum Fourier transform, $\mathcal{F}$, and its inverse, $\mathcal{F}^\dagger$. Therefore, in our framework, diagonalizing a circulant matrix $C$ satisfies
\begin{equation}
C = \mathcal{F}^\dagger \Lambda \mathcal{F}, \quad \text{i.e.,} \quad
\Lambda = \mathcal{F} C \mathcal{F}^\dagger.
\label{eq:diag_circ_matrix_qft}
\end{equation}

As a case study, we consider the eigenproblem for the circulant matrices $P_0$ and $P_1$ in Eq. \eqref{eq:P0_P1_matrix}. According to the notation introduced above for circulant matrices \eqref{eq:circ_matr} and their eigenvalues \eqref{eq:eigval_circ}, the matrix $P_0$ is defined by the elements $c_m = \delta_{m,1}$, its eigenvalues are $\lambda_m = \omega_N^{-m}$, and so its diagonal form is the matrix $\Omega^\dagger$ with $\Omega$ defined in Eq. \eqref{eq:Omega}. The matrix $P_1$ is defined by the elements $c_m = \delta_{m,N-1}$, its eigenvalues are $\lambda_m = \omega_N^{-(N-1)m} = \omega_N^{m}$, since $\omega_N^{-Nm} = \exp(-2 \pi i m )=1$, and so its diagonal form is the matrix $\Omega$.

\section[\appendixname~\thesection]{Optimization of the circuit for an initially localized walker}
\label{app:opt_init_state}
The initial state of a DTQW is typically assumed to be in the factorized form
\begin{equation}
\vert \psi_0 \rangle = \vert \phi_c \rangle \vert 0_p \rangle.
\label{eq:init_state}
\end{equation}

\noindent The coin is in a generic state $\vert \phi_c \rangle \in \mathcal{H}_c^{(2)}$, which can be realized by a proper unitary, $\vert \phi_c \rangle = U_c^{(0)} \vert 0_c \rangle$, starting from the default qubit state initialized to $\vert 0 \rangle$. Without loss of generality, the walker is localized in the $0$-th vertex. Indeed, any DTQW with an initially localized walker can be mapped onto a DTQW with a walker initialized in $\vert 0_p \rangle$ because of the equivalence, under cyclic permutation, of all the vertices in a cycle.
According to Eq. \eqref{eq:dtqw_T_matrix}, the first matrix acting on the initial state performs a QFT on the walker's position register. If the walker was in a generic position state, then the whole QFT-circuit would be required (see Fig. \ref{fig:our_qc}). However, given the initial state \eqref{eq:init_state} and Eq. \eqref{eq:QFT}, we observe that
\begin{equation}
{\rm QFT}:\,\vert 0 \rangle \mapsto \frac{1}{\sqrt{N}} \sum_{k=0}^{N-1} \vert k \rangle = \bigotimes_{m = 0}^{n-1} H_m \vert 0 \rangle^{\otimes n},
\label{eq:walk_init_state_Fourier}
\end{equation}

\noindent with $H_m$ the Hadamard gate 
$H = \frac{1}{\sqrt{2}}
\left( \begin{smallmatrix}
1 & 1\\1 & -1
\end{smallmatrix}\right)$
acting on the $m$-th qubit and $N=2^n$. Therefore, assuming the initial state \eqref{eq:init_state}, we simply need a layer of Hadamard gates. This is much more convenient than using the whole QFT-circuit, because Hadamard gates are one-qubit gates and can thus be executed in parallel. As a final remark, we point out that the above argument is unaffected from having used (I)QFT without SWAP gates in the circuit in Fig. \ref{fig:our_qc}, since $\tau (H \otimes \ldots \otimes H) \tau = H \otimes \ldots \otimes H$ with $\tau$ defined in Eq. \eqref{eq:swap_tau}.

\section[\appendixname~\thesection]{Analysis of the size of DTQW quantum circuits in different schemes}
\label{app:analysis_dtqw_qc}

In this appendix we provide details on the computation of the metrics---depth $\mathcal{D}$, number of one- and two-qubit gates, $\mathcal{N}^{(1)}$ and $\mathcal{N}^{(2)}$ respectively----of the quantum circuits implementing a DTQW on the $2^n$-cycle reported in Table \ref{tab:qc_metrics}. A few remarks:

(i) The ID-scheme here considered and the QFT-scheme are based on the same circuit design [Fig. \ref{fig:ID_qc}(d)], in which the coin gate (one-qubit gate) can not be executed in parallel to the CNOT gates, because it acts on their control qubit. Hence,  the coin gate has unit depth and prevents the simplification of CNOT gates when iteratively concatenating the single time-step quantum circuit to obtain the successive steps. Therefore, in Appendix \ref{subapp:ID_scheme}-\ref{subapp:QFT_scheme} the analysis focuses on the single time-step, because the size of a circuit implementing $t$ time-steps is $t$ times the size of that implementing the single time-step.

(ii) Controlled gates having different target qubits but the same control qubit (here is the coin) are executed in sequence (not in parallel) in actual quantum computers. This affects the depth of circuits and regards the CNOT gates in the ID- and QFT-scheme (Appendix \ref{subapp:ID_scheme} and \ref{subapp:QFT_scheme}, respectively) and the controlled-$R_k$ gates in our scheme (Appendix \ref{subapp:our_scheme}).

(iii) The (I)QFT on a $n$-qubit register (no SWAP) requires $n$ one-qubit gates (Hadamard), $n(n-1)/2$ two-qubit gates (controlled-$R_k$), and has depth $2n-1$ \cite{benenti_book}. This regards the analysis of the QFT-scheme (Appendix \ref{subapp:QFT_scheme}) and our scheme (Appendix \ref{subapp:our_scheme}).

\subsection{ID-scheme}
\label{subapp:ID_scheme}
We refer to the quantum circuit proposed in \cite{PhysRevA.79.052335} as the ID-scheme [Fig. \ref{fig:ID_qc}(a-c)], being based on increment and decrement gates. In particular, we assume the circuit to be designed as in Fig. \ref{fig:ID_qc}(d), because proved to be equivalent but more efficient (lower size and depth) than the standard ID-scheme in Fig. \ref{fig:ID_qc}(a-c) \cite{Shakeel2020}. This design requires only the increment gate (not controlled by the coin qubit) and a sequence of $2n$ CNOT gates [two-qubit gates, depth $2n$; see remark (ii)].
In this context, different realizations of the ID-scheme stem from different implementations of the generalized CNOT gates in the increment gate. In the following, we will refer to the generalized CNOT on $k \geq 3$ qubits as the $k$-Toffoli gate, i.e., a gate where one qubit is flipped conditional on the other $n_c = k-1$ (control) qubits being set to 1. A 3-Toffoli gate ($n_c = 2$) denotes the standard Toffoli gate. The increment gate on the $n$-qubit register [walker's position, see Fig. \ref{fig:ID_qc}(b)] requires $n-3$ $k$-Toffoli gates---with $k=4,\ldots,n$, i.e., $n_c=3,\ldots, n-1$---, one Toffoli gate ($k=3$, i.e., $n_c=2$), one CNOT ($n_c=1$, two-qubit gate), and one NOT ($n_c=0$, one-qubit gate), see Fig. \ref{fig:ID_qc}(b).

\textit{Implementation via linear-depth quantum circuit.---}A $k$-Toffoli gate, with $k \geq 3$, can be implemented using a linear-depth quantum circuit having $\mathcal{N}^{(1)}=0$, $\mathcal{N}^{(2)}=2k^2-6k+5$, and $\mathcal{D}=8k-20$ \cite{PhysRevA.87.062318}. The increment gate [Fig. \ref{fig:ID_qc}(b)] has $\mathcal{N}^{(1)}_{I} = 1$, $\mathcal{N}^{(2)}_{I} = 1 + \sum_{k=3}^{n}(2 k^2 - 6 k + 5)$, and $\mathcal{D}_{I} = 1 + 1 + \sum_{k=3}^{n}(8k-20)$. Therefore, the overall quantum circuit implementing the single time-step of the DTQW [Fig. \ref{fig:ID_qc}(d)] has
\begin{align}
& \mathcal{N}^{(1)}	= 1 + \mathcal{N}^{(1)}_{I}  = 2,\\
& \mathcal{N}^{(2)}	= 2 n + \mathcal{N}^{(2)}_{I} = \frac{1}{3}(2n^3-6n^2+13n-3),\\
& \mathcal{D}		= 1 + 2 n + \mathcal{D}_{I} = 4n^2-14n+19.
\end{align}

\textit{Implementation via ancilla qubits.---}A $k$-Toffoli gate, with $k \geq 4$, can be implemented using $\mathcal{N}^{(a)}=k-3$ ancilla qubits and $4(k-3)$ Toffoli gates \cite{PhysRevA.52.3457}. Accordingly, size and depth of this $k$-Toffoli gate are proportional to those of the Toffoli gate ($k=3$), which can be realized, e.g.,  with a linear-depth quantum circuit having $\mathcal{N}^{(1)}_{\rm Tof} = 0$, $\mathcal{N}^{(2)}_{\rm Tof} = 5$, and $\mathcal{D}_{\rm Tof} = 4$ (see previous paragraph) \cite{PhysRevA.87.062318}.
The increment gate [Fig. \ref{fig:ID_qc}(b)], in addition to one CNOT and one NOT, requires overall $1+\sum_{k=4}^{n}4(k-3)$ Toffoli gates (one standard Toffoli gate plus the decomposition of $k$-Toffoli gates with $k=4,\ldots,n$), so it has $\mathcal{N}^{(1)}_{I} = 1$, $\mathcal{N}^{(2)}_{I} = 1 + \left[ 1+\sum_{k=4}^{n}4(k-3) \right] \mathcal{N}^{(2)}_{\rm Tof}$ , and $\mathcal{D}_{I} = 1 + 1 + \left[ 1+\sum_{k=4}^{n}4(k-3) \right] \mathcal{D}_{\rm Tof} $. Therefore, the overall quantum circuit implementing the single time-step of the DTQW [Fig. \ref{fig:ID_qc}(d)] has
\begin{align}
& \mathcal{N}^{(1)}	= 1 + \mathcal{N}^{(1)}_{I}  = 2,\\
& \mathcal{N}^{(2)} = 2 n + \mathcal{N}^{(2)}_{I} = 10n^2-48n+66,\\
& \mathcal{D}		= 1 + 2  n + \mathcal{D}_{I} = 8n^2 - 38n + 55,\\
& \mathcal{N}^{(a)} = \max_{k\in\{4,\ldots,n\}}(k-3) = n-3.
\end{align}

\noindent The number of ancilla qubits is independent of the number of time-steps $t$ and is determined by the $k$-Toffoli gate having the largest number of controls, as the ancilla qubits can be reused among the $k$-Toffoli gates of the increment gate \cite{Shakeel2020}.

\textit{Implementation via rotations.---}To conclude this section, we mention another scheme, introduced in \cite{PhysRevA.103.022408}, which uses rotations around the basis states to implement the increment gate. However, as pointed out by the authors, the number of gates needed by this scheme increases exponentially faster than for the ID-scheme with ancillary qubits, although the latter scheme, as already discussed, quickly leads to a large workspace due to the number of required ancillary qubits which increases linearly with the number of control qubits. We do not delve into estimating $\mathcal{N}^{(1)}$, $\mathcal{N}^{(2)}$, and $\mathcal{D}$ in this scheme, so we refer the reader to \cite{PhysRevA.103.022408} for the detailed computation of the (total) number of gates involved.

\subsection{QFT-scheme}
\label{subapp:QFT_scheme}
We refer to the quantum circuit proposed in \cite{Shakeel2020} as the QFT-scheme (Fig. \ref{fig:QFT_qc}). This scheme, based on the circuit design in Fig. \ref{fig:ID_qc}(d), implements the increment gate by diagonalizing it via the QFT. In addition to the $2n$ CNOT gates [two-qubit gates, depth $2n$; see remark (ii)] and the (I)QFT [see remark (iii)], the circuit requires a ``layer'' of $n$ $R_k$ gates, which has unit depth (all the $n$ one-qubit gates are executed in parallel). Therefore, the overall quantum circuit implementing the single time-step of the DTQW (Fig. \ref{fig:QFT_qc}) has
\begin{align}
& \mathcal{N}^{(1)} = 1 + 2 n + n  = 3n +1,\\
& \mathcal{N}^{(2)} = 2 n + 2 \frac{n(n-1)}{2} = n^2 + n,\\
& \mathcal{D}		= 1 + 2 n + 2 (2n-1) + 1 = 6n.
\end{align}

\subsection{Present scheme}
\label{subapp:our_scheme}
The quantum circuit proposed in the present work is shown in Fig. \ref{fig:our_qc}. The (I)QFT is discussed in remark (iii). The ``layer'' of $n$ $R_k^\dagger$ gates and the coin gate has unit depth (all the $n+1$ one-qubit gates are executed in parallel). Then, the $n-1$ controlled-$R_k$ gates (two-qubit gates) amount to depth $n-1$ (sequential execution). Therefore, the overall quantum circuit implementing $t$ time-steps of the DTQW (Fig. \ref{fig:our_qc}) has
\begin{align}
& \mathcal{N}^{(1)}	= 2  n + t (n+1) = t(n +1)+2n,\\
& \mathcal{N}^{(2)}	= 2 \frac{n(n-1)}{2} + t (n - 1) = t (n-1) + n(n-1),\\
& \mathcal{D}			= 2 (2n-1) + t ( 1 + n - 1) = t n +2(2n-1).
\end{align}

\noindent These metrics comprise a fixed cost, due to the QFT and the IQFT, independent of the number $t$ of time-steps implemented, and a $t$-dependent cost. QFT and IQFT, unitary transformations, at the sides of the single time-step quantum circuit simplify each other when composing the circuit with itself.

\section[\appendixname~\thesection]{Transpilation of the proposed quantum circuit for the Hadamard DTQW on the 4- and 8-cycle}
\label{app:transpiled_dtqw_qc}
 In the present work, we consider the Hadamard DTQWs on the 4- and 8-cycle described in Sec. \ref{sec:H-DTQW}. For these DTQWs, we implement the circuit proposed in Fig. \ref{fig:our_qc} on the IBM quantum computer \texttt{ibm\_cairo} v.1.3.5, a 27-qubit Falcon r5.11 processor [qubit connectivity map in Fig. \ref{fig:ibm_cairo}(a)], whose native gates are
 \begin{align}
     &CNOT = \begin{pmatrix}
        1 & 0 & 0 & 0\\
        0 & 0 & 0 & 1\\
        0 & 0 & 1 & 0\\
        0 & 1 & 0 & 0\\
     \end{pmatrix},\quad
     I = \begin{pmatrix}
        1 & 0\\
        0 & 1\\
     \end{pmatrix} ,\quad
     R_z(\lambda) = \begin{pmatrix}
        e^{-i \frac{\lambda}{2}} & 0\\
        0 & e^{i \frac{\lambda}{2}}\\
     \end{pmatrix} ,\nonumber\\
    &\sqrt{X} = \frac{1}{2}\begin{pmatrix}
        1+i & 1-i\\
        1-i & 1+i\\
     \end{pmatrix} ,\quad
    X = \begin{pmatrix}
        0 & 1\\
        1 & 0\\
     \end{pmatrix} .
     \label{eq:native_gates_ibm_cairo}
\end{align}

\begin{figure}[H]
\includegraphics[width=\textwidth]{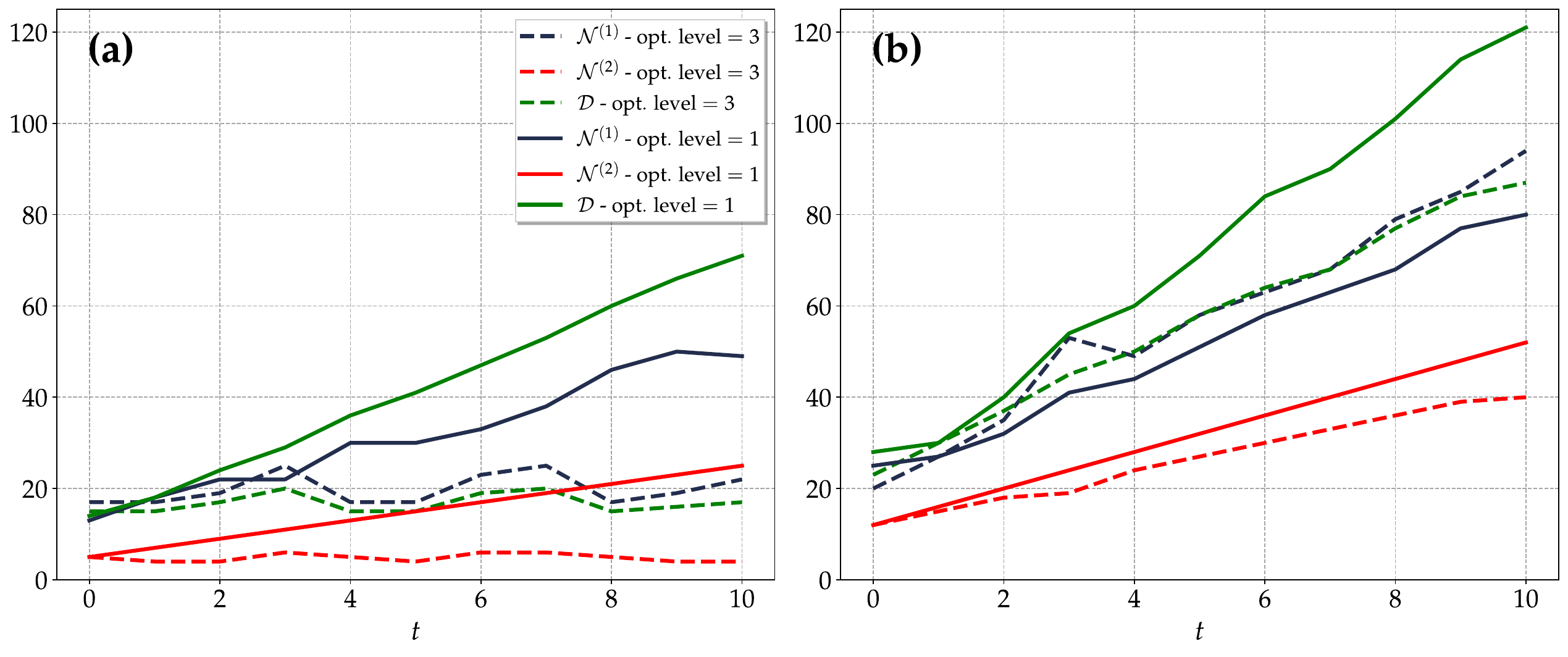}
\caption{Gate count in the quantum circuits proposed in the present work for the Hadamard DTQW on the \textbf{(a)} 4-cycle and \textbf{(b)} 8-cycle (Sec. \ref{sec:H-DTQW}): Depth $\mathcal{D}$ and number of one- and two-qubit gates, $\mathcal{N}^{(1)}$ and $\mathcal{N}^{(2)}$ respectively, as a function of time-steps $t$ with \texttt{optimization\_level=1,3} in transpilation (\texttt{ibm\_cairo}). The quantum circuits for $t=1$ are shown in Figs. \ref{fig:transpiled_2qubit} (4-cycle) and \ref{fig:transpiled_3qubit} (8-cycle).
\label{fig:gatecount_n2n3}}
\end{figure}

In Fig. \ref{fig:gatecount_n2n3} we show the gate count---number of one- and two-qubit gates---and depth of the quantum circuit implementing $t$ time-steps of the DTQW on the 4- and 8-cycle, panels (a) and (b) respectively, for $\texttt{optimization\_level=1,3}$ in transpilation. Consistently with Table \ref{tab:qc_metrics}, the general trend is linear in time for both the DTQWs on the 4- and 8-cycle with light optimization (\texttt{optimization\_level=1}) and only for the DTQW on the 8-cycle with the heaviest optimization (\texttt{optimization\_level=3}). In the particular case of the 4-cycle, for the latter optimization level the counts are basically independent of $t$ [Fig. \ref{fig:gatecount_n2n3}(a)].

\begin{figure}[H]
\begin{adjustwidth}{-\extralength}{0cm}
\includegraphics[width=\linewidth]{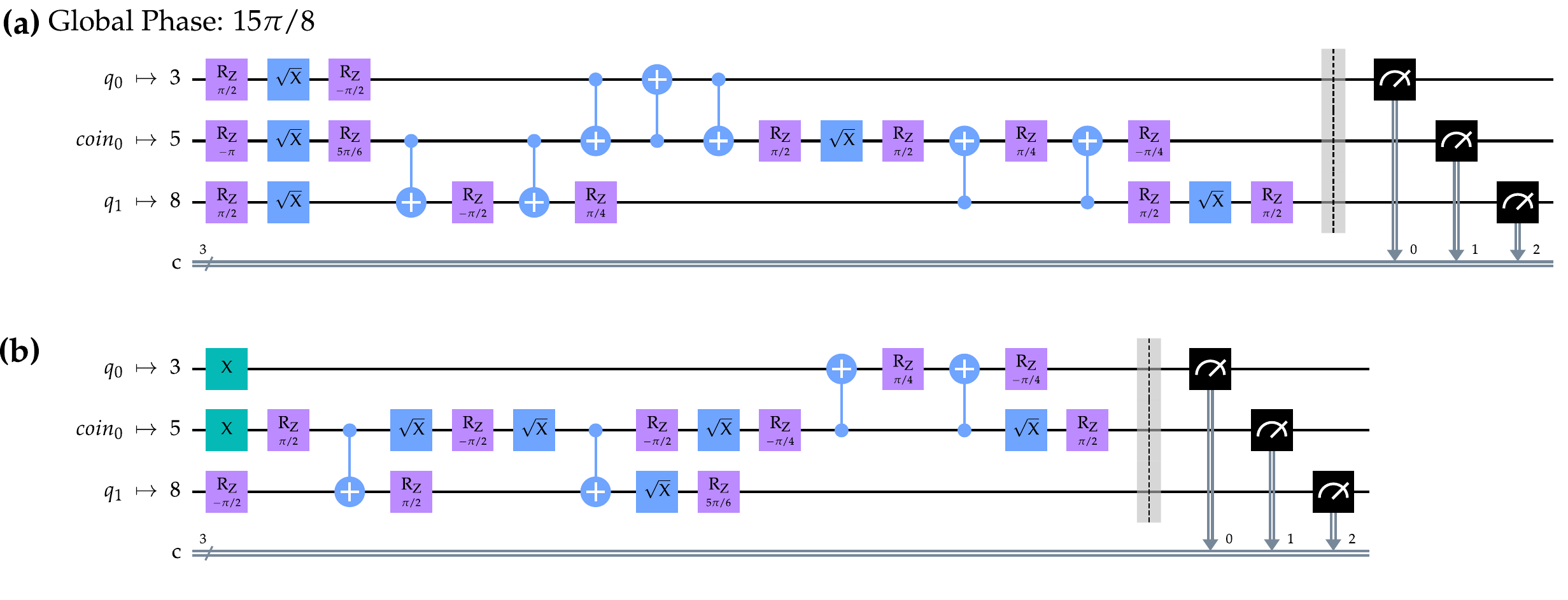}
\end{adjustwidth}
\caption{Quantum circuit proposed in the present work (Fig. \ref{fig:our_qc}) for one step of the Hadamard DTQW (Sec. \ref{sec:H-DTQW}) on the 4-cycle transpiled in \texttt{ibm\_cairo} with \textbf{(a)} \texttt{optimization\_level=1} and \textbf{(b)} \texttt{optimization\_level=3}. Each qubit is labelled by the corresponding index in the qubit connectivity map in Fig. \ref{fig:ibm_cairo}(a).
\label{fig:transpiled_2qubit}}
\end{figure}

\begin{figure}[H]
\begin{adjustwidth}{-\extralength}{0cm}
\includegraphics[width=\linewidth]{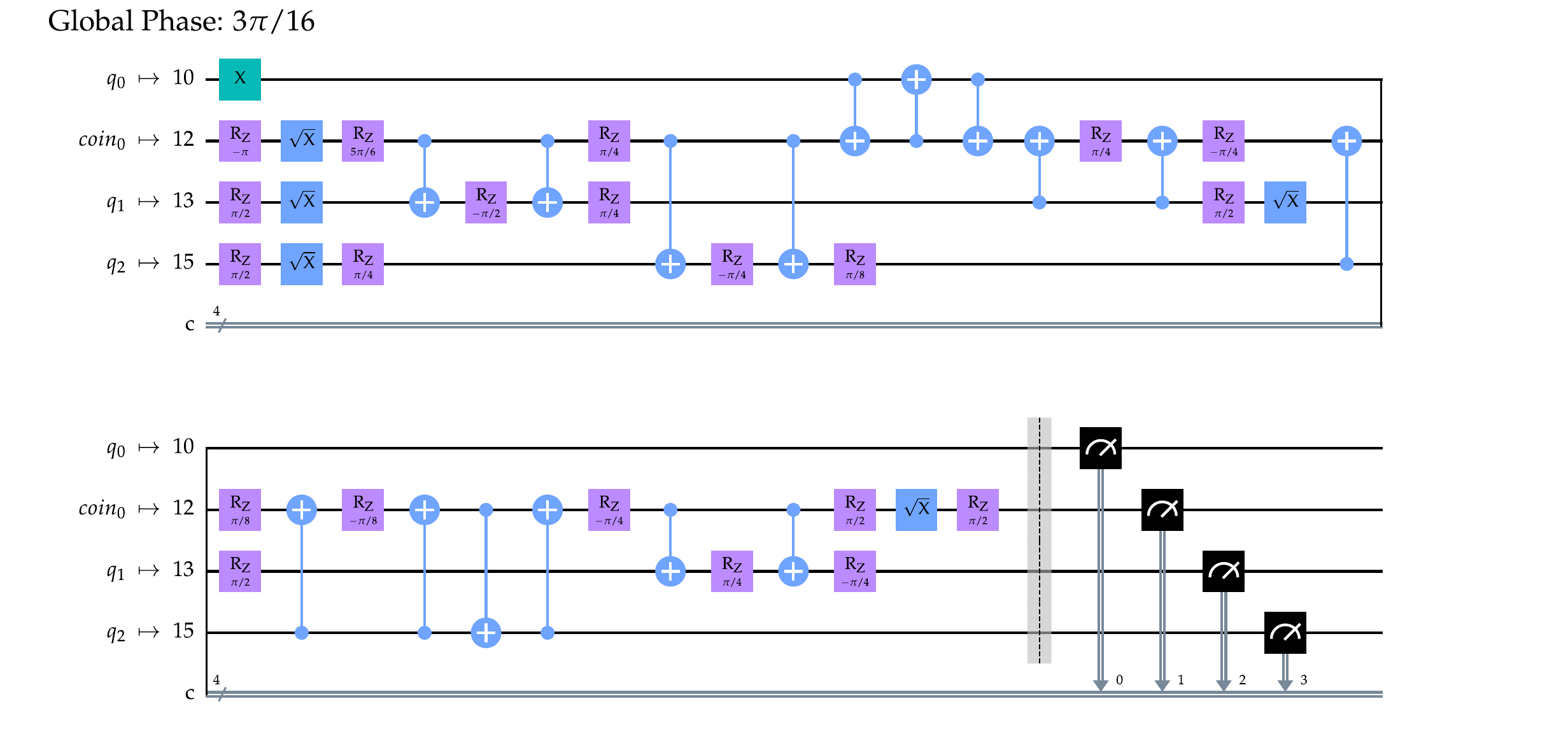}
\end{adjustwidth}
\caption{Same as in Fig. \ref{fig:transpiled_2qubit}, but for the 8-cycle with \texttt{optimization\_level=1}.
\label{fig:transpiled_3qubit}}
\end{figure}

The transpiled circuit implementing one step of the Hadamard DTQW is shown in Fig. \ref{fig:transpiled_2qubit} for the 4-cycle with \texttt{optimization\_level=1,3} and in Fig. \ref{fig:transpiled_3qubit} for the 8-cycle with \texttt{optimization\_level=1}. We recall that the initial state is given in Eq. \eqref{eq:psi0_ibm}, the initial QFT is implemented via Hadamard gates (Appendix \ref{app:opt_init_state}), and the Hadamard gate, which is not a native gate in this quantum processor, is transpiled into $H = R_z(\pi/2) \sqrt{X} R_z(\pi/2)$ [see Eq. \eqref{eq:Hadamard_coin} and Eq. \eqref{eq:native_gates_ibm_cairo}].

\begin{adjustwidth}{-\extralength}{0cm}

\reftitle{References}


\bibliography{dtqwibm_biblio}

%


\PublishersNote{}
\end{adjustwidth}
\end{document}